\documentclass[aip,superscriptaddress,showpacs,nofootinbib,
  floatfix,reprint]{revtex4-1}

\usepackage{dcolumn}
\usepackage{bm}
\usepackage{graphicx}
\usepackage{color}
\usepackage{hyperref}
\usepackage{latexsym}
\usepackage{amsthm}
\usepackage{amsmath}
\usepackage{amssymb}
\usepackage{verbatim}
\usepackage{setspace}
\usepackage{wasysym}
\usepackage{gensymb}

\usepackage[english]{babel}
\usepackage[utf8]{inputenc}

\newcommand{\cch}[1]{\left[#1\right]}

\newcommand{\prt}[1]{\left(#1\right)}
\newcommand{\aver}[1]{\left\langle #1 \right\rangle}

\DeclareRobustCommand{\uvec}[1]{{%
  \ifcsname uvec#1\endcsname
     \csname uvec#1\endcsname
   \else
    \bm{\hat{\mathbf{#1}}}%
   \fi
}}

\begin{document} 

 \title{Phase diagram and critical properties of a two-dimensional associating lattice gas 
}

\author{I. Ibagon}
\email{ibagon@fisica.ufmg.br}
\affiliation{Departamento de F\'isica, ICEx, Universidade Federal de Minas Gerais, C. P. 702, 30123-970, Belo Horizonte, Minas Gerais - Brazil}

\author{A. P. Furlan}
\email{apfurlan@fisica.ufmg.br}
\affiliation{Departamento de F\'isica, ICEx, Universidade Federal de Minas Gerais, C. P. 702, 30123-970, Belo Horizonte, Minas Gerais - Brazil}

\author{T. J. Oliveira}
\email{tiago@ufv.br}
\affiliation{Departamento de Física, Universidade Federal de Viçosa, 36570-900, Viçosa, Minas Gerais, Brazil}

\author{R. Dickman}
\email{dickman@fisica.ufmg.br}
\affiliation{Departamento de F\'isica and National Institute of Science and Technology for Complex Systems, ICEx, Universidade Federal de Minas Gerais, C. P. 702, 30123-970 Belo Horizonte, Minas Gerais - Brazil}

\date{\today}

\hypersetup{colorlinks=true, citecolor=blue,
  urlcolor=magenta,
  pdfcreator={pdflatex},
}

\begin{abstract} 
We revisit the associating lattice gas~(ALG) introduced by Henriques \textit{et al.} [PRE 71, 031504 (2005)] in its symmetric version. In this model, defined on the triangular lattice, interaction between 
molecules occupying nearest-neighbor sites depends on their relative orientation, mimicking the formation of hydrogen bonds in network-forming fluids. Although all previous studies of this model agree that it has a disordered fluid (DF), a low-density liquid (LDL) and a high-density liquid (HDL) phase, quite different forms have been reported for its phase diagram. Here, we present a thorough investigation of its phase behavior using both transfer matrix calculations and Monte Carlo (MC) simulations, along with finite-size scaling extrapolations. Results in striking agreement are found using these methods.  The critical point associated with the DF-HDL transition at full occupancy, identified by Furlan and coworkers [Phys.~Rev.~E~{\bf 100}, 022109 (2019)] is shown to be one terminus of a {\it critical line} separating these phases.
In opposition to previous simulation studies, we find that the transition between the DF and LDL phases is always discontinuous, similar to the LDL-HDL transition. The associated coexistence lines meet at the point where the DF-HDL critical line ends, making it critical-end-point. Overall, the form of the phase diagram observed in our simulations is very similar to that found in the exact solution of the model on a Husimi lattice. Our results confirm that, despite the existence of some waterlike anomalies in this model, it is unable to reproduce key features of the phase behavior of liquid water.
\end{abstract}
\maketitle

\section{Introduction}
\label{sec:Intro}

The importance water has to life on Earth is perhaps comparable to the degree of unusual behavior this liquid displays. In fact, when compared with other substances, liquid water presents a number of anomalies (such as in density, diffusivity and so on), which are believed to originate in the particular manner in which H$_2$O molecules interact, through hydrogen bonds forming a local, approximately tetrahedral network \cite{St13}. The three main scenarios proposed to understand water, {\it viz.} the stability-limit \cite{Sp76}, the singularity-free \cite{Sa96} and liquid-liquid critical point (LLCP) \cite{Po92} hypotheses,
are all consistent with the presence of such a network. The later scenarios are based on the fact that just as ice becomes glassy and can exist in low- and high-density amorphous phases \cite{Mi85}, the continuation of these phases ( increasing the temperature $T$ above that of spontaneous crystallization) may give rise to a low-density liquid (LDL) and a high-density liquid (HDL) phase \cite{St13,Ru10}. Although there is some indirect experimental evidence for the existence of the LLCP \cite{So00,Mi98,Wo18}, including some coming from water in nanopores \cite{Li05,Hsin06,Ma07,Ma09}, a definitive proof is still lacking \cite{So19}, mainly because the LLCP would exist in the metastable, deeply supercooled region, which is very difficult to access experimentally in bulk water.

In this context, much of this debate has been based on computational studies, with several realistic models for water confirming the LLCP hypothesis (see \cite{Je18} for a recent review). A number of simplified associating lattice gas (ALG) models, designed to possess two liquid phases, have also been introduced in two- \cite{BL70a,BL70b,Hu87,Pa99,Bu04,He05,Ba07} and three-dimensions \cite{Be72,Me82,La84,Be94,Ro96a,Ro96b,Pretti04,Pretti05,Sa93,Bo95,Sa96,Gi07}. These ALG models usually feature an attractive, orientation-dependent interaction between molecules occupying nearest neighbor (NN) sites --- mimicking the hydrogen bonding in network forming fluids ---, competing with a repulsive  excluded-volume
interaction. Phase diagrams displaying a LLCP have been reported for some of these models \cite{Bo95,Ro96a,Ro96b,Pretti04,Pretti05,He05,Gi07,Ba07}, though latter works on some these models have demonstrated a phase behavior without a LLCP \cite{He05,Gi07,Ba07}.

\begin{figure*}
    \centering
    \includegraphics[scale=.4]{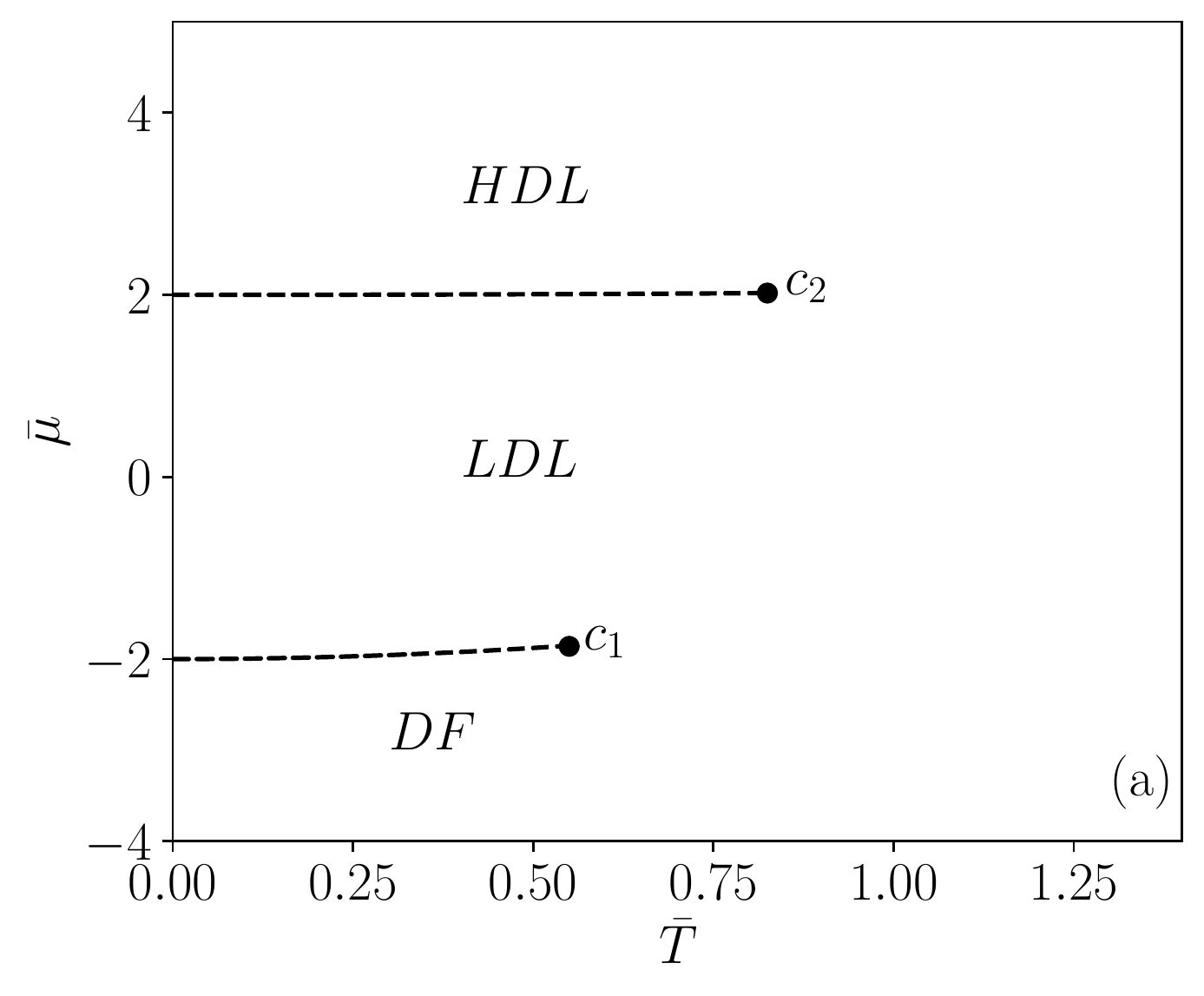}
    \includegraphics[scale=.4]{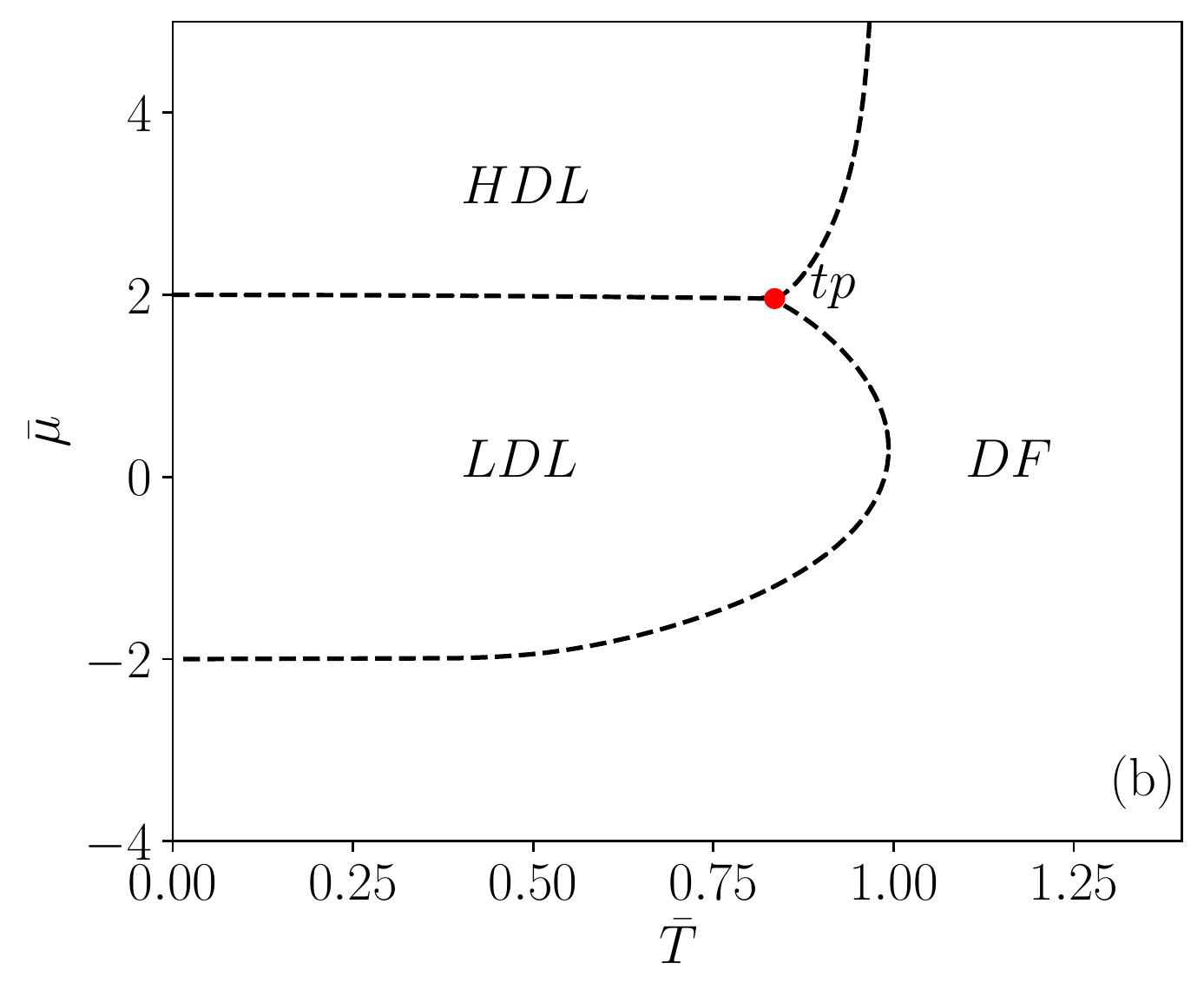}
    \includegraphics[scale=.4]{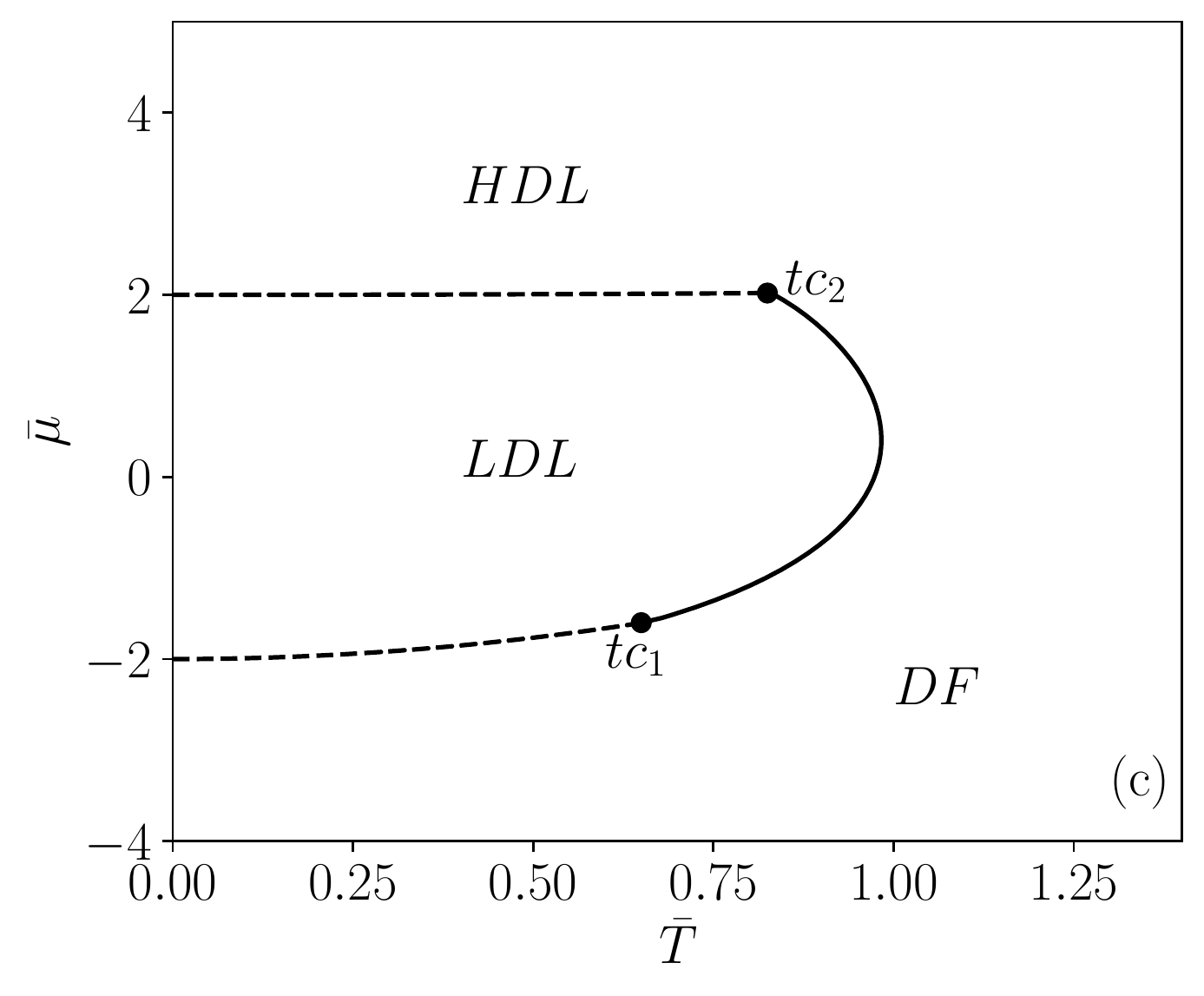}
    \caption{Qualitative phase diagrams for the symmetric H\&B model in the (reduced) $\bar{\mu}-\bar{T}$ plane adapted from references \cite{Ba07,Ol10,Fu15} in panels (a), (b), and (c), respectively. Solid and dashed lines represent continuous and discontinuous phase transitions, respectively. The coexistence points at $T=0$ are exact, while the loci of the critical points $c_1$ and $c_2$, of the triple point $tp$, and of the tricritical points $tc_1$ and $tc_2$ are those found numerically in the references cited above. }
\label{fig:reffigs}
\end{figure*}

An example of this is the symmetric version of the Besseling and Lyklema model \cite{Be94}, firstly analyzed by Girardi \textit{et al.} \cite{Gi07}. This model is defined on the body-centered cubic lattice, where each molecule can form up to four hydrogen bonds with NNs, similar to H$_2$O. While these bonds lower their energy, nonbonding molecules at NN sites experience a repulsive ``van der Waals'' interaction. 
Beyond the GAS-liquid coexistence line ending at a critical point, a LDL-HDL coexistence line ending at a second critical point was also reported in \cite{Gi07}, yielding a phase diagram qualitatively similar to the one in Fig. \ref{fig:reffigs}(a). (As an aside, we remark that the phase commonly called GAS in the literature related to these models will be regarded as a \textit{disordered fluid} (DF) here, since it may extend to large chemical potentials and so, high densities.) Subsequent studies of this ALG \cite{Bu08,Bu09,Sz10}, notwithstanding, revealed a much more complex phase behavior, with the LLCP [$c_2$ in Fig. \ref{fig:reffigs}(a)] being actually a tricritical point, while $c_1$ is disputed as being a critical-end-point \cite{Bu08,Bu09} or a bicritical point \cite{Sz10}. These works also uncovered a critical line separating the HDL and DF phases \cite{Bu08,Bu09,Sz10}.

A similar sequence of events marks the history of the ALG model introduced by Henriques and Barbosa (H\&B) \cite{He05} 
which is a nonsymmetric two-dimensional (2D) version of the 3D model analyzed by Girardi \textit{et al.} \cite{Gi07}. The H\&B model is defined on the triangular lattice, in which each molecule has four bonding arms (two donors and two acceptors of protons), beyond two inert arms separated by 180\degree. In their original Monte Carlo analysis, H\&B \cite{He05} reported a phase diagram analogous to that in Fig. \ref{fig:reffigs}(a), with two coexistence lines and two critical points. Further work on this model indicated, however, a different and richer phase diagram, with two additional continuous transition lines: DF-LDL and DF-HDL~\cite{Sz09}. In this phase diagram, the DF-LDL coexistence line ends at a tricritical point where it meets the DF-LDL continuous line; and the LDL-HDL coexistence line ends at a bicritical point where it meets the continuous DF-LDL and DF-HDL lines~\cite{Sz09}.

Several modifications of the H\&B model have been proposed~\cite{Ba07,Fu16,Fu15}. Of particular interest here is the symmetric version introduced by Balladares {\it et al.}~\cite{Ba07}, in which no distinction is made between donor and acceptor arms. The literature on this symmetric ALG is also marked by controversy. For example, the original MC simulations of Ref. \cite{Ba07} again suggested the phase behavior of Fig.~\ref{fig:reffigs}(a), consistent with the LLCP hypothesis. The semi-analytical mean-field solution of this simplified version of the H\&B model on a Husimi lattice built with hexagonal plaquettes \cite{Ol10} revealed a qualitatively different phase diagram, with three coexistence lines (DF-HDL, DF-LDL and LDL-HDL) meeting at a triple point ($tp$), as shown in Fig. \ref{fig:reffigs}(b). Later MC simulations \cite{Fu15} supported the phase behavior of Fig. \ref{fig:reffigs}(c), with two coexistence lines (DF-LDL and LDL-HDL) ending at two tricritical points ($tc_1$ and $tc_2$), which are linked by a critical DF-LDL line. No transition between the HDL phase and the disordered fluid was reported in Ref. \cite{Fu15}.

More recently, this symmetric H\&B model was investigated in the limit of full occupancy (corresponding to a chemical potential $\mu \to \infty$) using MC simulations and semi-analytical solutions on Husimi lattices \cite{Fu19}. The simulations yielded a continuous transition between the HDL and the DF phase at a (reduced) temperature $\bar{T}_c=0.9525(5)$, with critical exponents and Binder cumulant in good agreement with those of the three-state Potts model in 2D \cite{Wu82}. On the other hand, the Husimi lattice solutions furnish a discontinuous transition, as a continuation of the DF-HDL coexistence line found for finite $\mu $\cite{Ol10} to the limit of $\mu \rightarrow \infty$. All these results raise important questions on the phase behavior of this model: \textit{i)} Does it indeed exhibit a LLCP as in Fig. \ref{fig:reffigs}(a)? \textit{ii)} What is the true nature of the DF-LDL transition? Is it discontinuous as in Fig. \ref{fig:reffigs}(b) or does it have a tricritical point and a continuous part like in Fig. \ref{fig:reffigs}(c)? \textit{iii)} Assuming that a transition line does indeed exist separating the DF and HDL phases: Is it always discontinuous but ends at a critical point in the limit $\mu \to \infty$? Or does it start discontinuous and change to continuous at some finite $\mu$? Or, even, is the entire line continuous? 

In this work we address these points through an extensive numerical study of this ALG model using both transfer-matrix analysis and MC simulations in the grand-canonical ensemble. Consistent results are obtained from both methods, demonstrating that the correct topology of the phase diagram is that of Fig. \ref{fig:reffigs}(b). However, in contrast with the mean-field result \cite{Ol10} shown in this figure, we find that the transition between the DF and HDL phases is always continuous and the critical DF-HDL line ends at a critical-end-point, where it meets the DF-LDL and LDL-HDL coexistence lines. The critical properties of the continuous transition line --- an interesting problem scarcely tackled in the literature of lattice models for water (see \cite{Fu19} for a discussion) --- are also investigated and an intriguing behavior is found, with some quantities presenting a close agreement with the three-state Potts class, while others have a large deviation from it.

The paper is organized as follows. In Sec. \ref{sec:model}, we present the investigated model. The transfer-matrix method and the associated results are discussed in Sec.~\ref{sec:TM}. A similar thing is done in Sec. \ref{sec:MC} for the Monte Carlo simulations. The comparison of the outcomes from both methods and our concluding remarks are presented in Sec. \ref{sec:Conc}.

\section{Model}
\label{sec:model}

\begin{figure}
    \centering
    \includegraphics[scale=1]{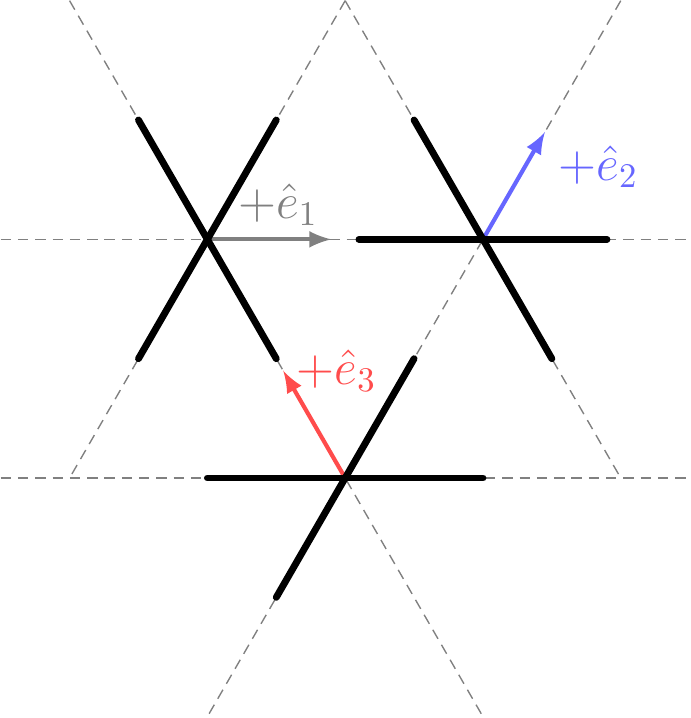}
    
    \caption{Possible orientational states ${\bm \eta} \equiv \hat{e}_{k}$, with $k=1,2$ and 3. Bonding arms are represented by thick black lines and the generators of the triangular lattice $\hat{e}_1$, $\hat{e}_2$ and $\hat{e}_3$ are indicated by tiny coloured arrows. For this configuration no hydrogen bond is formed between these molecules. }
    \label{fig:defstates}
\end{figure}

We study the symmetric version of the Henriques and Barbosa (H\&B) \cite{He05} associating lattice gas~(ALG) model, investigated in Refs. \cite{Ba07,Ol10,Fu15,Fu19}. The model is defined on the triangular lattice (coordination number $z=6$), where each site $i$ can be either empty ($\sigma_i=0$) or occupied by a single molecule ($\sigma_i=1$). A chemical potential $\mu$ is associated to each particle in the system, since we will work in the grand canonical ensemble. Whenever two molecules occupy nearest-neighbor (NN) sites, they experience a repulsive ``van der Walls" interaction, gaining an energy $\varepsilon > 0$. In addition, two NN molecules can form a hydrogen bond, which decreases their energy by a factor $-\gamma<0$, provided that their bonding arms point toward each other. In fact, a key feature of the model analyzed here (and of other ALGs as well) is the existence of orientational degrees of freedom and related interactions introduced by bonding arms. In our case, each molecule has four bonding arms and two inert (nonbonding) arms, as shown in Fig. \ref{fig:defstates}. The four bonding arms are all equivalent and always interact with those of NN molecules in the same manner, i.e., there is no distinction between donors and acceptors (of protons) here, in opposition to the original (non-symmetric) H\&B model \cite{He05}. The inert arms of a given molecule are in diametrically opposed lattice edges (forming an angle 180\degree), which results in three orientational states for each molecule (see Fig. \ref{fig:defstates}.) Following Ref.~\cite{Fu19}, we may conveniently define an orientational variable, ${\bm \eta}_i$, pointing along one of the nonbonding directions of the molecule at site $i$. In this fashion, it is simply given by the generators of the triangular lattice, $\hat{e}_{k}$ with $k=1,2$ and 3, which are also depicted in Fig. \ref{fig:defstates}. Namely, ${\bm \eta}_i \equiv \hat{e}_{k}$, where $\hat{e}_1=\uvec{i}$, $\hat{e}_2=+\frac{1}{2}\uvec{i}+\frac{\sqrt{3}}{2}\uvec{j}$ and $
  \hat{e}_3=-\frac{1}{2}\uvec{i}+\frac{\sqrt{3}}{2}\uvec{j}$. 
  
With these definitions, we may write the Hamiltonian of this ALG model as,
\begin{equation}\label{eq:hamil}
  {\cal H}\prt{{\bm \eta},{\bm r}}= \sum_{\aver{i,j}}
  \sigma_i\sigma_j\cch{\varepsilon + u_{ij}}- \mu \sum_i\sigma_i,
\end{equation}
where ${\bm r} = {\bm r}_i - {\bm r}_j$ is a unit vector in the set $\left\lbrace \pm\hat{e}_1, \pm\hat{e}_2, \pm\hat{e}_3 \right\rbrace$. The second sum gives the total number of particles, $N$, while the first sum runs over all NN pairs $(i,j)$, whose orientational interaction is enconded in variable $u_{ij}$, given by~\cite{Fu19}
\begin{equation}\label{eq:uij}
    u_{ij} = -\gamma\prt{\frac{4}{3}}^2
  \cch{1-\prt{{\bm \eta}_i\cdot {\bm r}}^2}
  \cch{1-\prt{{\bm \eta}_j\cdot {\bm r}}^2},
\end{equation}
Note that whenever ${\bm r}$ is parallel or antiparallel to ${\bm \eta}_i$ and/or ${\bm \eta}_j$, molecules $(i,j)$ do not form a hydrogen bond and $u_{i,j}=0$; otherwise $u_{i,j}=-\gamma$. Following most of the previous studies of the ALG model \cite{Ba07,Ol10,Fu15}, we will use the parameter $\varepsilon$ as the energy scale and set $\gamma/\varepsilon=2$. In this way, the net interaction between two NN molecules forming (not forming) a hydrogen bond is attractive (repulsive). We therefore have as free parameters the dimensionless temperature $\bar{T} = k_B T/\varepsilon$ and chemical potential $\bar{\mu} = \mu/\varepsilon$. We remark that in Ref. \cite{Fu19} the temperature (called $\tau$ there) was defined in units of $\gamma$, so that $\tau = \bar{T}/2$.

\begin{figure}
    \centering
    \includegraphics[scale=0.75]{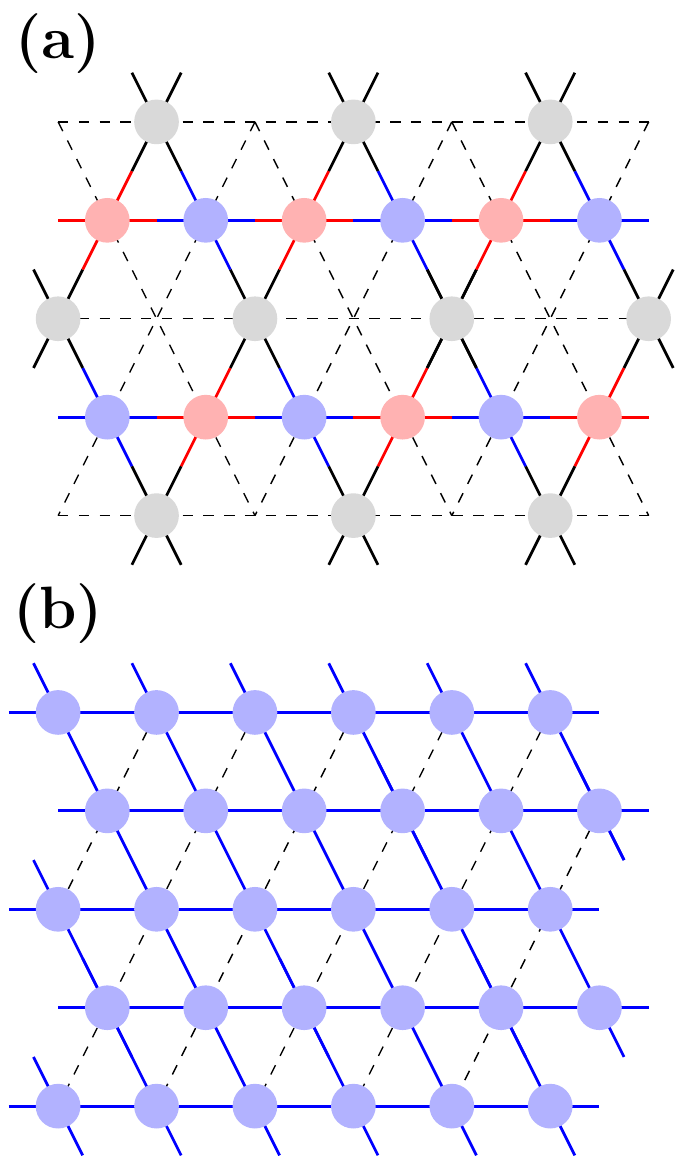}
    
    \caption{Illustration of a ground state configuration of the (a) low-density liquid (LDL) phase and (b) high-density liquid (HDL) phase. Different colors represent molecules in distinct orientational states. The bonding arms are indicated by thick lines. }
    \label{fig:picphases}
\end{figure}

As already discussed in the Introduction [Sec. \ref{sec:Intro}], from previous studies of the ALG model~\cite{Ba07,Ol10,Fu15}, one knows that its phase diagram exhibits three phases: disordered fluid (DF), low-density liquid~(LDL) and high-density liquid~(HDL). In the ground state ($\bar{T}=0$), the DF phase has particle density $\rho=0$ (the reason it is usually called GAS) and is the most stable phase for $\bar{\mu} < -2$. For $-2<\bar{\mu}<2$ the LDL phase is the one with the lowest free energy. In this phase, one of the four sublattices is empty while each of the other three is occupied by molecules in one of the orientational states, such that each molecule participates in four hydrogen bonds~[see Fig.~\ref{fig:picphases} (a)]. Hence, the LDL phase has $\rho=3/4$ and a fourfold degeneracy. For $\bar{\mu} > 2$ the HDL phase is the most stable, with the lattice fully occupied ($\rho=1$) and all inert arms aligned along the same direction, as illustrated in Fig.~\ref{fig:picphases}(b). The HDL phase is threefold degenerate. It is quite simple to show that the DF-LDL and LDL-HDL phases coexist at $\bar{\mu}=-2$ and $\bar{\mu}=2$, respectively, when $\bar{T}=0$ \cite{Ba07,Ol10}.

\section{Transfer-matrix analysis}
\label{sec:TM}

\subsection{Preliminaries}
\label{sec:metTM}

\begin{figure}[t]
  \includegraphics[width=8.5cm]{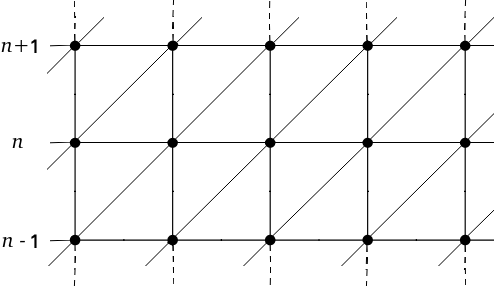}
  \caption{Illustration of the approach to the triangular lattice used in our transfer-matrix calculations. A strip of width $L=5$ is shown.}
  \label{fig:strip}
\end{figure}

We define the transfer matrix, $\tilde{M}_L$, of the model on strips of infinite length and finite width $L$. As illustrated in Fig. \ref{fig:strip}, we consider square lattice strips with diagonal edges, with periodic boundary conditions in the finite (horizontal) direction. As usual, each possible state for a given row of the strip will be associated with a row and a column of $\tilde{M}_L$. Since each lattice site can be in four different states, we have a total of $4^L$ states for a strip of width $L$, so that $\tilde{M}_L$ has dimension $4^L \times 4^L$. In order to investigate the full-occupancy case ($\bar{\mu} \rightarrow \infty$), where only three states are possible for each site, we can work with reduced matrices of dimension $3^L \times 3^L$. In any case, the element $(i,j)$ of $\tilde{M}_L$ is obtained by setting the row $n$ in state $i$, the row $n+1$ in state $j$ and determining the statistical weight of this configuration. Then, $\tilde{M}_L(i,j) = \exp[(N_p \bar{\mu} + 2 N_{hb} - N_{vdW})/2 \bar{T}]$, where $N_p$ is the total number of particles on both rows (note that we are dividing this by 2), and $N_{hb}$ and $N_{vdW}$ counts the \textit{effective} number of hydrogen bonds and van der Waals interactions (i.e, first neighbors) in these rows, respectively. By ``effective'', we mean that we are counting once (twice) interactions along the horizontal (diagonal and vertical) direction and, then, dividing them by 2. In this way, the ``internal'' interactions, along the vertical and diagonal directions, are counted once, while each one in the horizontal contribute with $1/2$ in each state, since they appear in two states.

It is hard to deal directly with $\tilde{M}_L$, because all of its elements are non-null. Therefore, in order to save computer RAM memory and speed up the calculations, we replace the procedure of adding an entire row, by the procedure of adding a single elementary cell (formed by squares with a diagonal edge). Working in this way, we have to deal with $L+1$ matrices, $\tilde{T}_L^{(s)}$, so that $\tilde{M}_L = \tilde{T}_L^{(L)} \cdot \tilde{T}_L^{(L-1)}\cdots\tilde{T}_L^{(0)}$. Here, $\tilde{T}_L^{(0)}$ creates a new site in row $n+1$ above a complete row $n$; the matrices $\tilde{T}_L^{(s)}$, with $1 \leqslant s \leqslant L-1$, add new elementary cells to row $n+1$; and finally $\tilde{T}_L^{(L)}$ completes the row $n+1$ by imposing the appropriated (periodic here) boundary conditions (see, e.g, Refs. \cite{guo02,Ol15} for more details). The advantage of this method relies on the fact that the matrices $\tilde{T}_L^{(s)}$ are very sparse. In fact, the total number of non-null terms summing over all matrices $\tilde{T}_L^{(s)}$ is only $(16L-8)4^L$.

\begin{table*}[bth]
\begin{ruledtabular}
\begin{tabular}{llllllll}
      $L$ &    $\bar{T}_{c,1}$  &      $X_{c,1}$      &    $c^{(1)}$      & &  $\bar{T}_{c,2}$   &      $X_{c,2}$      &    $c^{(2)}$   \\
      \hline  
       3  &   0.9630812528   &  0.1153561805   &  0.9439481004     & &  0.9831360600   &  0.1951137102   &  0.9400187288  \\ 
       4  &   0.9581778321   &  0.1122361525   &  0.9019637901     & &  0.9689256192   &  0.1810700192   &  0.8996045252  \\
       5  &   0.9546313471   &  0.1090076877   &  0.8651449784     & &  0.9611962232   &  0.1708572156   &  0.8639745823  \\ 
       6  &   0.9530663416   &  0.1071998513   &  0.8438815963     & &  0.9575263053   &  0.1648349316   &  0.8433113197  \\
       7  &   0.9524498263   &  0.1063326346   &  0.8322698047     & &  0.9557272984   &  0.1612950160   &  0.8319934887  \\
       8  &   0.9522229606   &  0.1059533630   &  0.8256760684     & &  0.9547690621   &  0.1590849673   &  0.8255533747  \\
       9  &   0.9521529738   &  0.1058159572   &  0.8216678338     & &  0.9542109914   &  0.1576018062   &  0.8216323836  \\
      10  &   0.9521470966   &  0.1058006641   &  0.8190595772     & &  0.9538600720   &  0.1565417468   &  0.8190774634  \\
      11  &   0.9521675217   &  0.1058493262   &  0.8172623591     & &  0.9536256433   &  0.1557459461   &  0.8173147797  \\
      12  &   0.9521981542   &  0.1059325169   &  0.8159665891     & &  0.9534614919   &  0.1551257762   &  0.8160423453  \\
      13  &   0.9522318614   &  0.1060345131   &  0.8149983146     & &  0.9533421966   &  0.1546283056   &  0.8150903408  \\
      14  &   0.9522654200   &  0.1061464695   &  0.8142535202     & &  0.9532528474   &  0.1542199764   &  0.8143571526  \\
      15  &   0.9522974007   &  0.1062632003   &  0.8136666597     & &  0.9531842311   &  0.1538784508   &  0.8137787074  \\
\end{tabular}
\caption{Finite-size estimates of the pseudocritical temperature $\bar{T}_{c,k}$ and scaling dimension $X_{c,k}$ from the condition $X_k(L-1,\bar{T}_{c,k})=X_k(L+1,\bar{T}_{c,k})$ for $k=1$ (left) and $k=2$ (right). The central charges $c^{(k)}$, calculated at $\bar{T}_{c,k}(L)$, are also shown.}
\label{tab:MTfull1}
\end{ruledtabular}
\end{table*}

If $\Lambda_L^{(k)}$ denotes the eigenvalues of $\tilde{M}_L$, for given values of $\bar{\mu}$ and $\bar{T}$, with $\Lambda_L^{(0)} > |\Lambda_L^{(1)}| > |\Lambda_L^{(2)}| > \ldots$, then, the (dimensionless) negative free energy density is
\begin{equation}
 f(L) = \frac{\zeta }{L} \ln \Lambda_L^{(0)},
\end{equation}
where $\zeta$ is a geometric factor given by the thickness of the added row when its circumference is $L$, which in our case is $\zeta = 2/\sqrt{3}$. The correlation length $\xi_k(L)$ associated with the $k$th subleading eigenvalue is given by
\begin{equation}
 \xi_k^{-1}(L) = \zeta \ln \frac{\Lambda_L^{(0)}}{|\Lambda_L^{(k)}|}.
 \label{eq:corr_len}
\end{equation}
From these correlation lengths, we may defined the scaled gaps $X_k(L) = L/[2\pi \xi_k(L)]$. Note that all these quantities obviously depend also on $\bar{\mu}$ and $\bar{T}$. Close to a critical point ($\bar{T}_c$ or $\bar{\mu}_c$), from finite-size scaling \cite{fisher71}, we expect that
\begin{equation}
 X_k(L;\bar{T},\bar{\mu}) = X_k^* + a_{t} t L^{y_t} + a_{1} L^{y_1} + a_{2} L^{y_2} + \ldots,
 \label{eq:FSgap}
\end{equation}
where $y_t>0$ is the temperature exponent, $y_i<0$ ($i=1,2,\ldots$) are irrelevant exponents, and $t=\bar{T}-\bar{T}_c$ if $\bar{\mu}$ is kept fixed or $t=\bar{\mu}-\bar{\mu}_c$ when $\bar{T}$ is the fixed parameter.

In conformally invariant systems, the critical free energy density is expected to follow
\begin{equation}
 f(L) = f_{\infty} + \frac{\pi c}{6 L^2} + p_1 L^{z_1}+ \ldots,
 \label{eq:FSfreeenergy}
\end{equation}
where $c$ is the conformal anomaly \cite{blote86,*affleck86} and $z_1$ is a negative exponent.

\subsection{The full-occupancy limit}

Let us start analyzing the limiting case $\bar{\mu} \rightarrow \infty$, where all lattice sites are occupied by molecules. Here, we can work with smaller transfer matrices, as noticed above, what allow us to investigate strip widths up to $L=16$. As recently demonstrated by some of us, in such a limit the system displays only the DF and HDL phases, separated by a critical point in the class of the two-dimensional (2D) 3-state Potts model \cite{Fu19}.

If a critical point exists at some critical temperature $\bar{T}_c^*$, from Eq. \ref{eq:FSgap}, one may expect that curves of the scaled gaps $X_k(L-\ell;\bar{T})$ and $X_k(L+\ell;\bar{T})$ versus $\bar{T}$ (e.g., with $\ell = 1$) cross at pseudocritical temperatures $\bar{T}_{c,k}(L)$, which shall converge to $\bar{T}_c^*$ as $L \rightarrow \infty$. Therefore, $\bar{T}_{c,k}(L)$ can be determined from the condition $X_k(L-1,\bar{T}_{c,k})=X_k(L+1,\bar{T}_{c,k}) \equiv X_{c,k}(L)$, from which a pseudocritical estimate [$X_{c,k}(L)$] of the asymptotic scaling dimension $X_k^*$ is also obtained. The values of these quantities for $k=1$ and $k=2$ are shown in Tab. \ref{tab:MTfull1}. The first thing to note there is that $\bar{T}_{c,1}$ has a non-monotonic behavior, initially decreasing and then increasing with $L$. This prevents any reliable extrapolation of this temperature. The convergence of $\bar{T}_{c,2}$, on the other hand, is monotonic and, so, we will focus on it to estimate $\bar{T}_c^*$. 

In general, from finite-size scaling, one expects that
\begin{equation}
 \bar{T}_{c,k}(L) = \bar{T}_c^* + b_1 L^{-\omega_1} + b_2 L^{-\omega_2} + \ldots,
 \label{eq:FStau}
\end{equation}
whose exponents are related to those in Eq. \ref{eq:FSgap} as $\omega_1=y_t-y_1$ and $\omega_2=y_t-y_2$. Then, by considering $b_i=0$ for $i \geqslant 2$ and performing three-point (3-pt) extrapolations of the values of $\bar{T}_{c,2}$ in Tab. \ref{tab:MTfull1} (with $\bar{T}_c^*$, $b_1$ and $\omega_1$ as unknowns in Eq. \ref{eq:FStau}), we find the temperatures displayed in Tab. \ref{tab:MTfullext}. Note that for the largest $L$'s they are varying at the fifth decimal place. A second round of 3-pt extrapolations [now of the extrapolated values in Tab. \ref{tab:MTfullext}] yields $0.952806 \leqslant \bar{T}_c^* \leqslant 0.952818$, for the four largest sets of widths, without any clear tendency to increase or decrease. Hence, this gives $\bar{T}_c^* = 0.952812(6)$. This value differs by only $0.02$\% from the one estimated from MC simulations in Ref. \cite{Fu19}: $\bar{T}_c^* \approx 0.9526$. 

The resulting exponents $\omega_1$ from the first 3-pt extrapolations are also displayed in Tab. \ref{tab:MTfullext}. Three-point extrapolations of them, assuming again a power-law correction, return values close (and converging toward) to $\omega_1=2$ for the largest sets of widths. We remark that, according to Queiroz \cite{Queiroz00}, the irrelevant exponents in Eq. \ref{eq:FSgap} are given by $y_1=-4/5$ and $y_2=-8/5$ for the 3-state Potts model, whose critical thermal exponent is $y_t=6/5$. This means that $\omega_1=2$ and $\omega_2=14/5=2.8$ in this case. By assuming that $\omega_1=2$ and $b_i=0$, for $i \geqslant 3$ in Eq. \ref{eq:FStau}, we estimate the exponents $\omega_2$ for different $L$'s from 4-pt extrapolations; and subsequent 3-pt extrapolations of them yield values close to $14/5$ for the largest $L$'s. So, the correction exponents of the ALG model in the full lattice regime are consistent with the ones for the 3-state Potts model.

\begin{table}[t]
\begin{ruledtabular}
\begin{tabular}{llll}
      $L$    &   $\bar{T}_{c}^*$  &    $\omega_1$   &      $c$      \\
      \hline  
       4     &   0.9397502     &   1.3793263   &     --        \\
       5     &   0.9516285     &   2.6536683   &   0.7857046   \\
       6     &   0.9529237     &   3.2158880   &   0.8086578   \\
       7     &   0.9530776     &   3.3614896   &   0.8119752   \\
       8     &   0.9530316     &   3.2893957   &   0.8121525   \\
       9     &   0.9529681     &   3.1489286   &   0.8117281   \\
      10     &   0.9529211     &   3.0137856   &   0.8112940   \\
      11     &   0.9528904     &   2.9040695   &   0.8109670   \\
      12     &   0.9528705     &   2.8173407   &   0.8107358   \\
      13     &   0.9528570     &   2.7468393   &   0.8105667   \\
      14     &   0.9528474     &   2.6875514   &   0.8104331   \\

\end{tabular}
\caption{Extrapolated temperatures, $\bar{T}_{c}^*$, and corresponding correction exponents, $\omega_1$, from 3-pt extrapolations of the values of $\bar{T}_{c,2}$ displayed in Tab. \ref{tab:MTfull1}, considering Eq. \ref{eq:FStau} and sets of sizes ($L-1,L,L+1$). The resulting central charges, $c$, from analogous extrapolations of the values of $c^{(2)}$ in Tab. \ref{tab:MTfull1} are also shown.}
\label{tab:MTfullext}
\end{ruledtabular}
\end{table}

The finite-size amplitudes $X_{c,1}$ and $X_{c,2}$ in Tab. \ref{tab:MTfull1}, on the other hand, do not agree with the ones expected for the 2D 3-state Potts class. In fact, for conformally invariant models in this class, the asymptotic values of these quantities are $X_{1}^* = 2 - y_h = 2/15$ and $X_{2}^* = 2 - y_t = 4/5$. However, our values of $X_{c,1}$ for the largest $L$'s are $\sim 20$\% smaller than $2/15$, though they display a non-monotonic convergence and are increasing for large $L$'s. So, it may be the case that for very large strip widths they will attain the expected Potts value. For $X_{c,2}$, our values are smaller than $4/5$ by a factor $\sim 5$ and are still decreasing, without any indication of changing in this behavior. Actually, the values in Tab. \ref{tab:MTfull1} extrapolate to $X_{2}^* \approx 0.15$, suggesting that $y_t \approx 1.85$ and then $\nu=1/y_t \approx 0.54$. Thereby, we can be either dealing with a system displaying very slow convergences in these quantities (what seems strange in the case of $X_{c,2}$...) or the ALG is not a conformally invariant model or it belongs to a different universality class.

The central charge is somewhat consistent with the 3-state Potts class, for which $c=4/5$. In fact, the values of $c$, calculated as $c(L) = 3[f(L-1)-f(L+1)](L-1)^2(L+1)^2/2\pi L$ at the pseudocritical temperatures $\bar{T}_{c,1}$ and $\bar{T}_{c,2}$, are shown in Tab. \ref{tab:MTfull1}; and they approximate $4/5$ as the system size increases. However, 3-pt extrapolations of such estimates (assuming power-law corrections) return $c \gtrsim 0.81$ [see Tab. \ref{tab:MTfullext}]. Further 3-pt extrapolations of the extrapolated values in Tab. \ref{tab:MTfullext}, considering the largest $L$'s available, do not improve this, once they yield $c \approx 0.809$, which differs by $\sim 1$\% from the Potts value.

\begin{table}[t]
\begin{ruledtabular}
\begin{tabular}{lllll}
      $L$    &      $f$        &     $X_1$     &    $X_2$     &     $c$       \\
      \hline  
       2     &    5.029729     &    0.112107   &   0.180682   &      --       \\
       3     &    4.960323     &    0.109445   &   0.168819   &   0.944204    \\
       4     &    4.937032     &    0.107661   &   0.161966   &   0.902030    \\
       5     &    4.926737     &    0.106952   &   0.158348   &   0.865111    \\
       6     &    4.921304     &    0.106774   &   0.156322   &   0.843869    \\
       7     &    4.918080     &    0.106823   &   0.155076   &   0.832298    \\
       8     &    4.916008     &    0.106965   &   0.154240   &   0.825738    \\
       9     &    4.914595     &    0.107143   &   0.153641   &   0.821754    \\
      10     &    4.913587     &    0.107334   &   0.153189   &   0.819162    \\
      11     &    4.912844     &    0.107525   &   0.152836   &   0.817377    \\
      12     &    4.912280     &    0.107713   &   0.152551   &   0.816090    \\
      13     &    4.911841     &    0.107895   &   0.152317   &   0.815052    \\
      14     &    4.911494     &    0.108069   &   0.152121   &   0.814293    \\
      15     &    4.911213     &    0.108237   &   0.151953   &   0.813804    \\
      16     &    4.910984     &    0.108397   &   0.151809   &      --

\end{tabular}
\caption{Critical free energies, $f$, and scaled gaps $X_1$ and $X_2$ calculated at the critical temperature $\bar{T}_c^* = 0.952812$. The central charges, $c$, estimated from these free energies are also shown.}
\label{tab:MTfullCP}
\end{ruledtabular}
\end{table}


Table \ref{tab:MTfullCP} shows the scaled gaps $X_1$ and $X_2$ calculated at the critical point $\bar{T}_c^* = 0.952812$. However, these quantities are very similar to those in Tab. \ref{tab:MTfull1}, obtained at the pseudocritical temperatures. The critical free energies at $\bar{T}_c^* = 0.952812$, as well as the central charges obtained from them (calculated as indicated above) are also depicted in Tab. \ref{tab:MTfullCP}. Once again, the values of $c$ present an appreciable difference from $c=4/5$ and extrapolations of them have deviations similar to those observed above.

\subsection{The DF-HDL transition line}
\label{subsec:DF-HDL}

\begin{figure}[tb]
  \includegraphics[width=8.5cm]{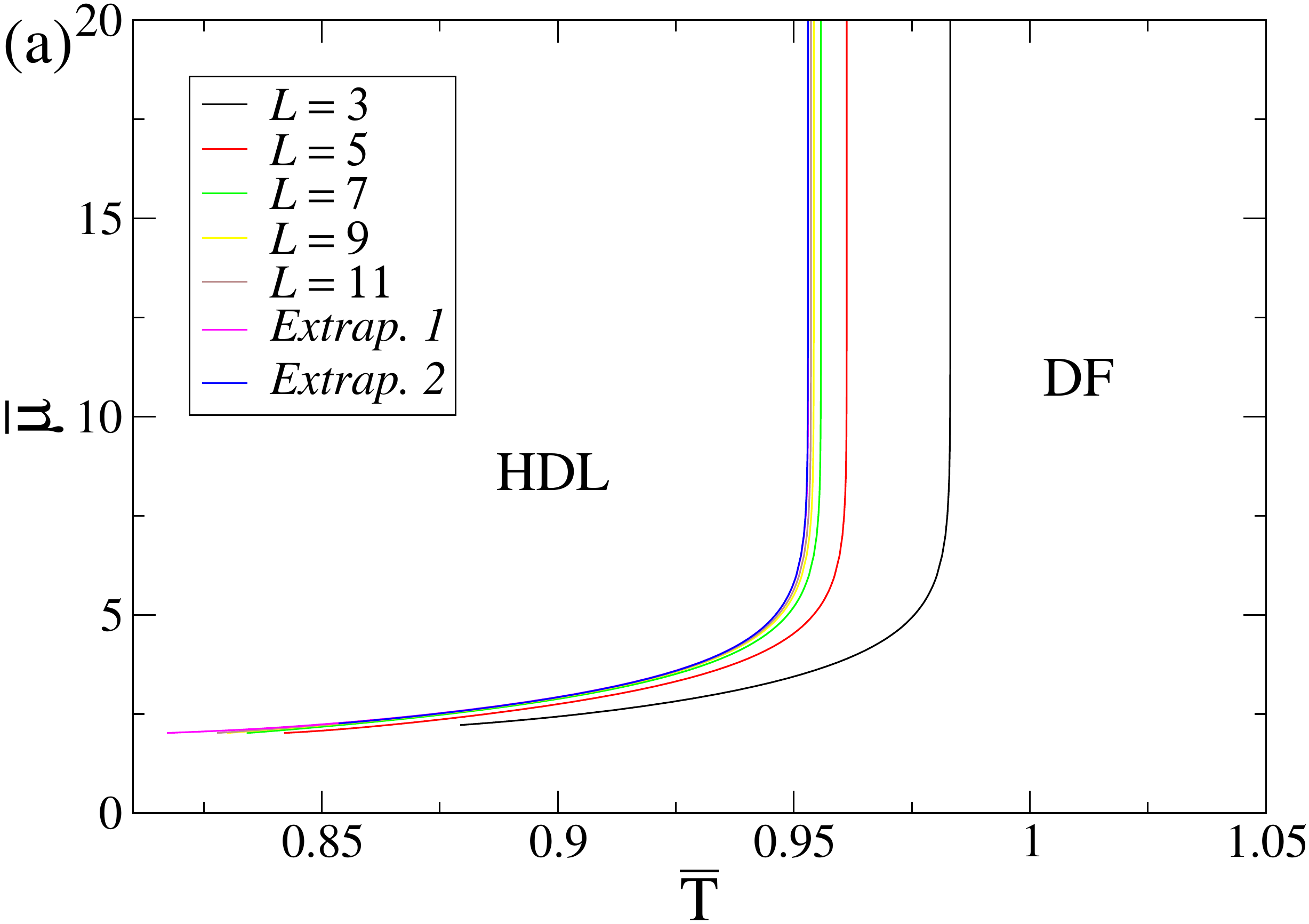}
  \includegraphics[width=8.5cm]{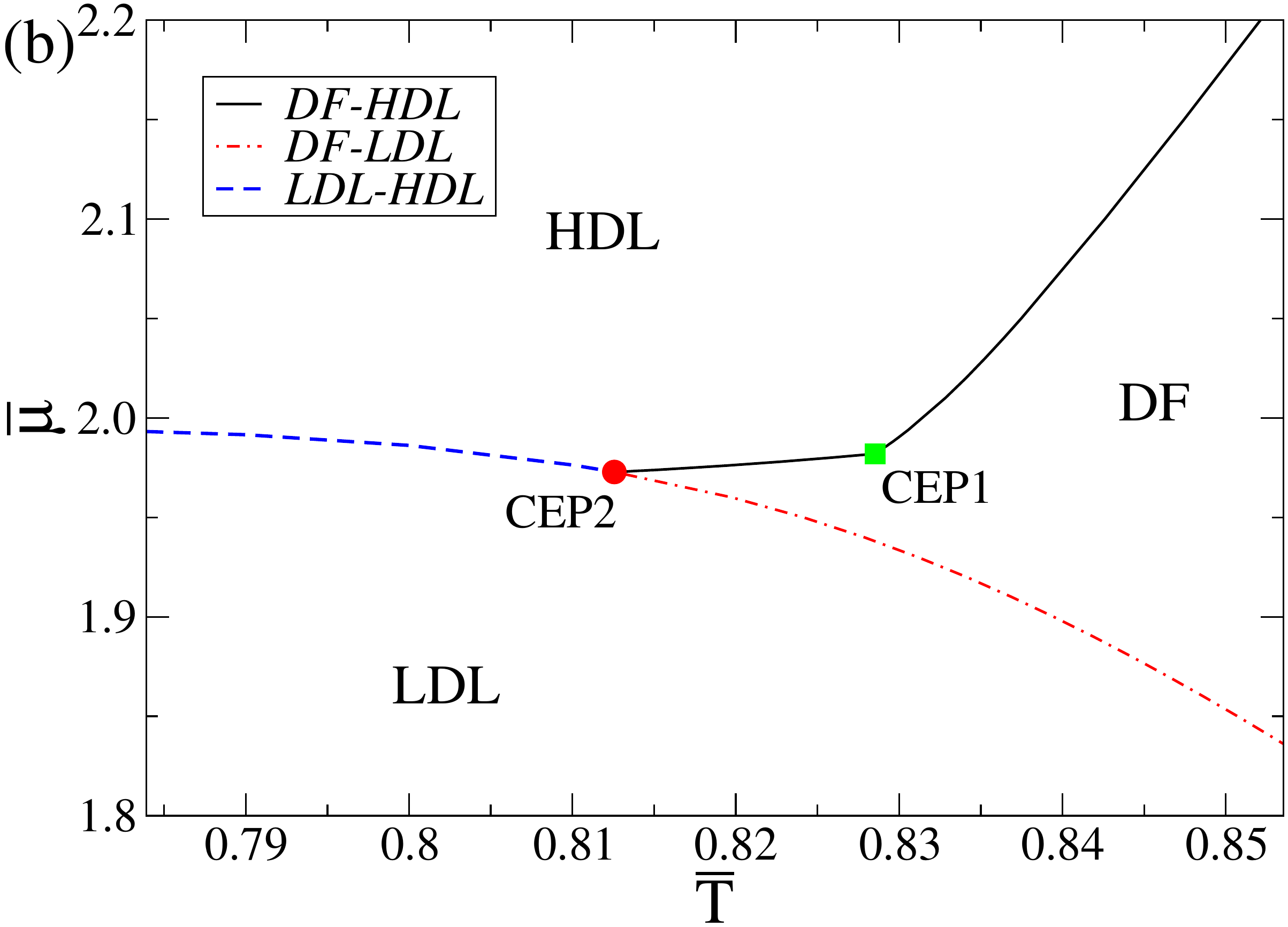}
  \caption{(Color online) a) Pseudocritical lines separating the DF and HDL phases, obtained from $X_2(L-1,\bar{T}_c^*)=X_2(L+1,\bar{T}_c^*)$ for fixed values of $\bar{\mu}$, for the indicated $L$'s. The outcomes from 3-pt extrapolations of these lines considering the sets $(5,7,9)$ [Extrap. 1] and $(7,9,11)$ [Extrap. 2] are also shown. b) Detail of the region where the DF-HDL, DF-LDL and LDL-HDL transition lines, for $L=7$, meet. The green square and the red circle indicate the two estimates for the critical-end-point.}
  \label{fig:GASHDL}
\end{figure}

Employing the same analysis from the previous subsection, but for finite $\bar{\mu}$, we find that the critical DF-HDL transition extends to chemical potentials down to $\bar{\mu} \sim 2$, giving rise to a DF-HDL transition line, as found in the mean-field solutions of the model on Husimi lattices \cite{Ol10,Fu19}. Given the non-monotonic convergence observed above in the critical points estimated from $X_1$, we will focus here only on those calculated from $X_2$. The pseudocritical transition lines obtained from the condition $X_2(L-1,\bar{T}_c^*)=X_2(L+1,\bar{T}_c^*)$, for fixed values of $\bar{\mu}$, are shown in Fig. \ref{fig:GASHDL}(a), for $L \leqslant 11$. (Here, it is quite hard to go beyond $L=12$, once the TMs have $4^L \times 4^L$ terms. Actually, most of our results will be limited to $L\leqslant 10$.) It is important to notice that such lines never cross each other, indicating that the DF-HDL transition is always continuous.

Three-point extrapolations of the pseudocritical lines --- assuming the finite-size correction of Eq. \ref{eq:FStau} for fixed $\bar{\mu}$, with $a_{i}=0$ for $i>1$, and considering the sets of widths $(5,7,9)$ and $(7,9,11)$ --- yield lines whose values of $\bar{T}_c^*$ always differ by less than $0.2$\%. This corresponds to differences (and error bars) at most at the third decimal place, which is accurate enough for our purposes.

Interestingly, the pseudocritical lines have a corner at $\bar{\mu} \approx 2$, which may be seen as their end-point. An example of this is shown in Fig. \ref{fig:GASHDL}(b), in the line for $L=7$. As it will be demonstrated below, the critical DF-HDL line ends at a critical end-point (CEP). The coordinates ($\bar{T}_{cep,1},\bar{\mu}_{cep,1}$) of the CEP, assuming that it is located at the corner, are depicted in Tab. \ref{tab:CEP}. Since the DF-HDL transition lines continue below these corners and finally meet the DF-LDL and LDL-HDL transition lines [see Fig. \ref{fig:GASHDL}(b)], one obtains another set of estimates ($\bar{T}_{cep,2},\bar{\mu}_{cep,2}$) for the CEP. As shows Tab. \ref{tab:CEP}, in this last case the temperatures oscillate around $\bar{T}_{cep,2} = 0.81266$ (considering the three largest $L$'s). The temperatures of the corners seem to converge toward this value as $L$ increases; though a 3-pt extrapolation of them (following Eq. \ref{eq:FStau}, considering the largest $L$'s) returns a slightly larger value: $\bar{T}_{cep,1} = 0.81681$. This deviation, of $\sim 0.5$\%, is certainly due to the small widths considered here, so that it is reasonable to regard $\bar{T}_{cep} = 0.815(3)$ as our estimate for the CEP temperature. This value is close, but a bit smaller than $\bar{T}_{c/tc} = 0.825$, which was found in previous MC simulations of the ALG model and reported as a critical point in \cite{Ba07} and as a tricritical point in \cite{Fu15}. Our estimate is also smaller than the value found for the triple point ($\bar{T}_{tp}=0.835$) in the Husimi lattice solution of the model \cite{Ol10}.

\begin{table}[t]
\begin{ruledtabular}
\begin{tabular}{lllll}
$L$      & $\bar{T}_{cep,1}$ & $\bar{\mu}_{cep,1}$    & $\bar{T}_{cep,2}$    & $\bar{\mu}_{cep,2}$  \\ 
\hline
3        &   0.884217     &    2.264032            &    0.801210       &  2.010048    \\
5        &   0.841799     &    2.018722            &    0.812741       &  1.982581    \\
7        &   0.828528     &    1.981947            &    0.812563       &  1.972774    \\
9        &   0.823467     &    1.970668            &    0.812692       &  1.968213    \\
\end{tabular}
\caption{Estimates for the critical-end-point from the corner in the DF-HDL transition lines (leftmost values) and from the point where such lines meet the other transition lines (rightmost values).}
\label{tab:CEP}
\end{ruledtabular}
\end{table}

The 3-pt extrapolations of the values of $\bar{\mu}_{cep,1}(L)$ and $\bar{\mu}_{cep,2}(L)$ for the largest $L$'s, assuming power-law corrections similarly to Eq. \ref{eq:FStau}, yield $\bar{\mu}_{cep,1}=1.9589$ and $\bar{\mu}_{cep,2}=1.9605$, respectively, indicating that $\bar{\mu}_{cep}=1.9597(8)$. This result agrees with the mean-field value found in Ref. \cite{Ol10} for the triple point ($\bar{\mu}_{tp}=1.959$) and differs by $\approx 3$\% from the critical point reported in Ref. \cite{Ba07} ($\bar{\mu}_c = 2.02$). 
\begin{figure}[b]
  \includegraphics[width=8.5cm]{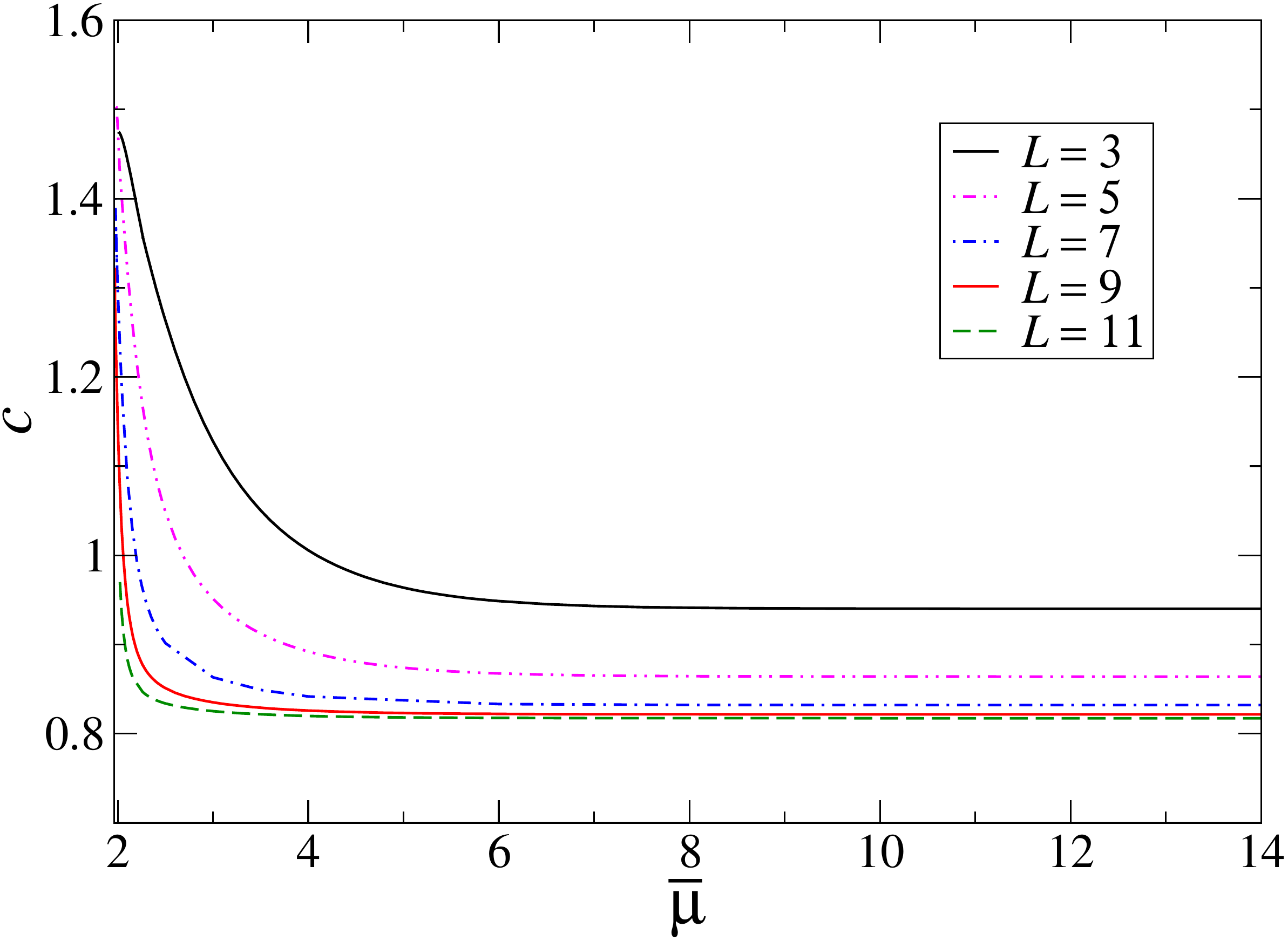}
  \caption{(Color online) Central charge $c$ versus $\bar{\mu}$ calculated at the pseudocritical transition lines from Fig. \ref{fig:GASHDL}(a), for the indicated widths.}
  \label{fig:GASHDLcc}
\end{figure}

The central charges calculated along the pseudocritical transition lines from Fig. \ref{fig:GASHDL}(a) are shown in Fig. \ref{fig:GASHDLcc} as a function $\bar{\mu}$. We may note that $c(L)$ is almost independent of the chemical potential for $\bar{\mu}_c \gtrsim 10$ (similarly to $\bar{T}_c$), having values very close to those of $c^{(2)}$ in Tab. \ref{tab:MTfull1} for $\bar{\mu} \rightarrow \infty$. When $\bar{\mu}$ approximates $\bar{\mu}_{cep}$, however, one sees that $c$ passes to increase and to display stronger finite-size corrections. The large $L$ is the close the approximately constant part of the lines get to $\bar{\mu}_{cep}$ before deviating upward. This strongly suggests that for very large $L$'s one must have $c \approx 0.8$ in the entire DF-HDL line, indicating that it belongs to the 3-state Potts class.

\subsection{The LDL-HDL transition line}
\label{subsec:LDL-HDL}
\begin{figure}[!htb]
  \includegraphics[width=8.5cm]{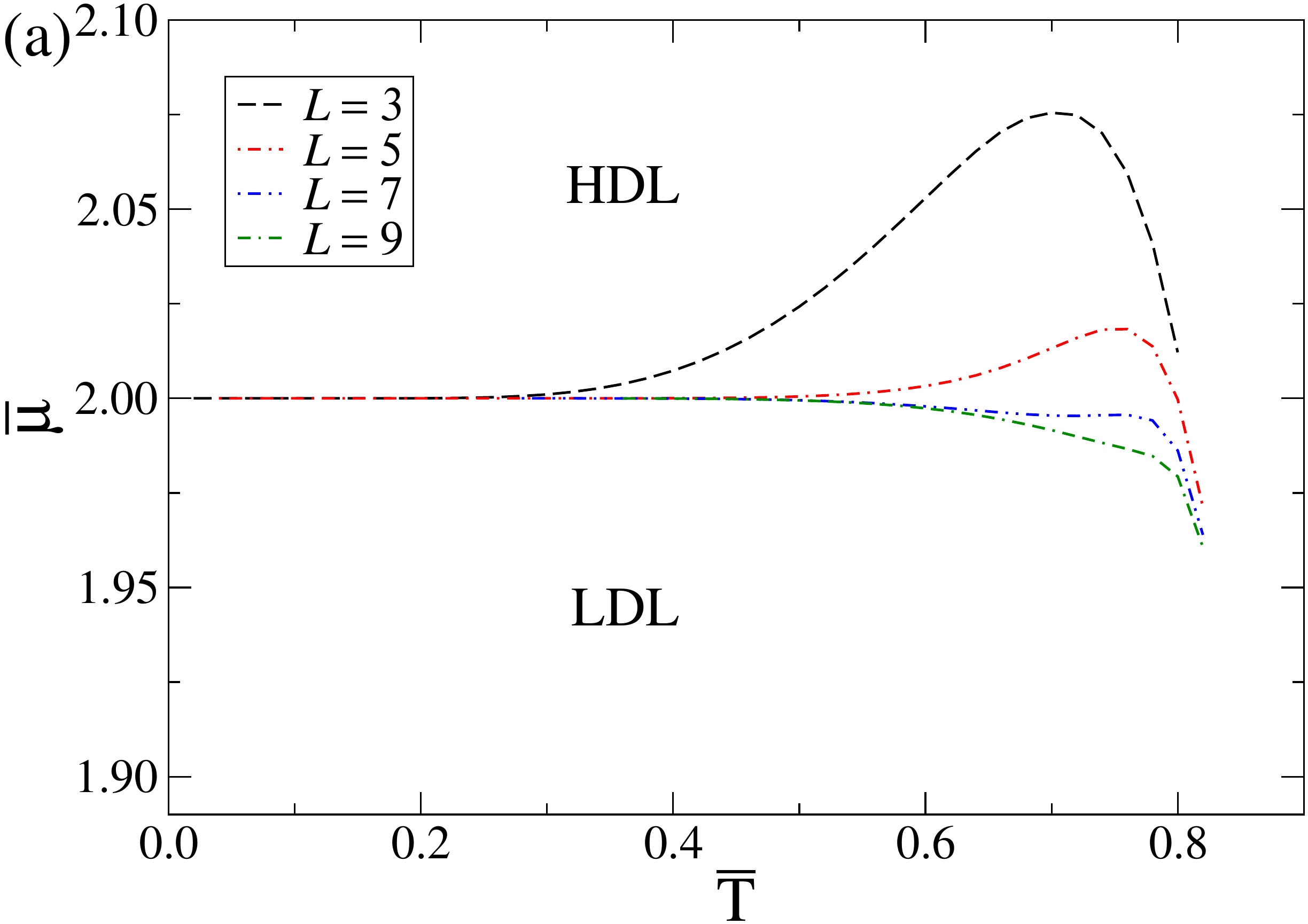}
  \includegraphics[width=8.5cm]{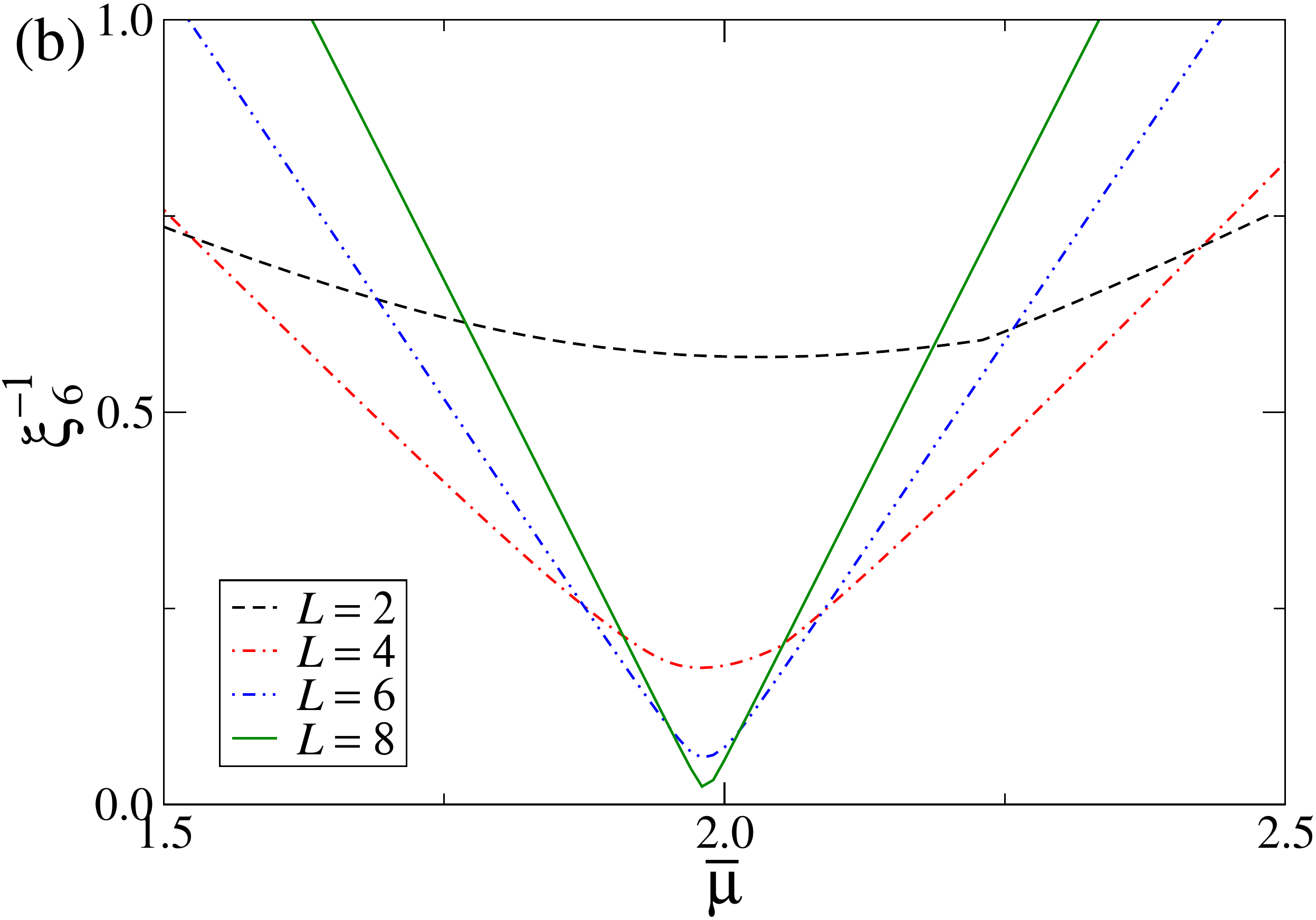}
  \includegraphics[width=8.5cm]{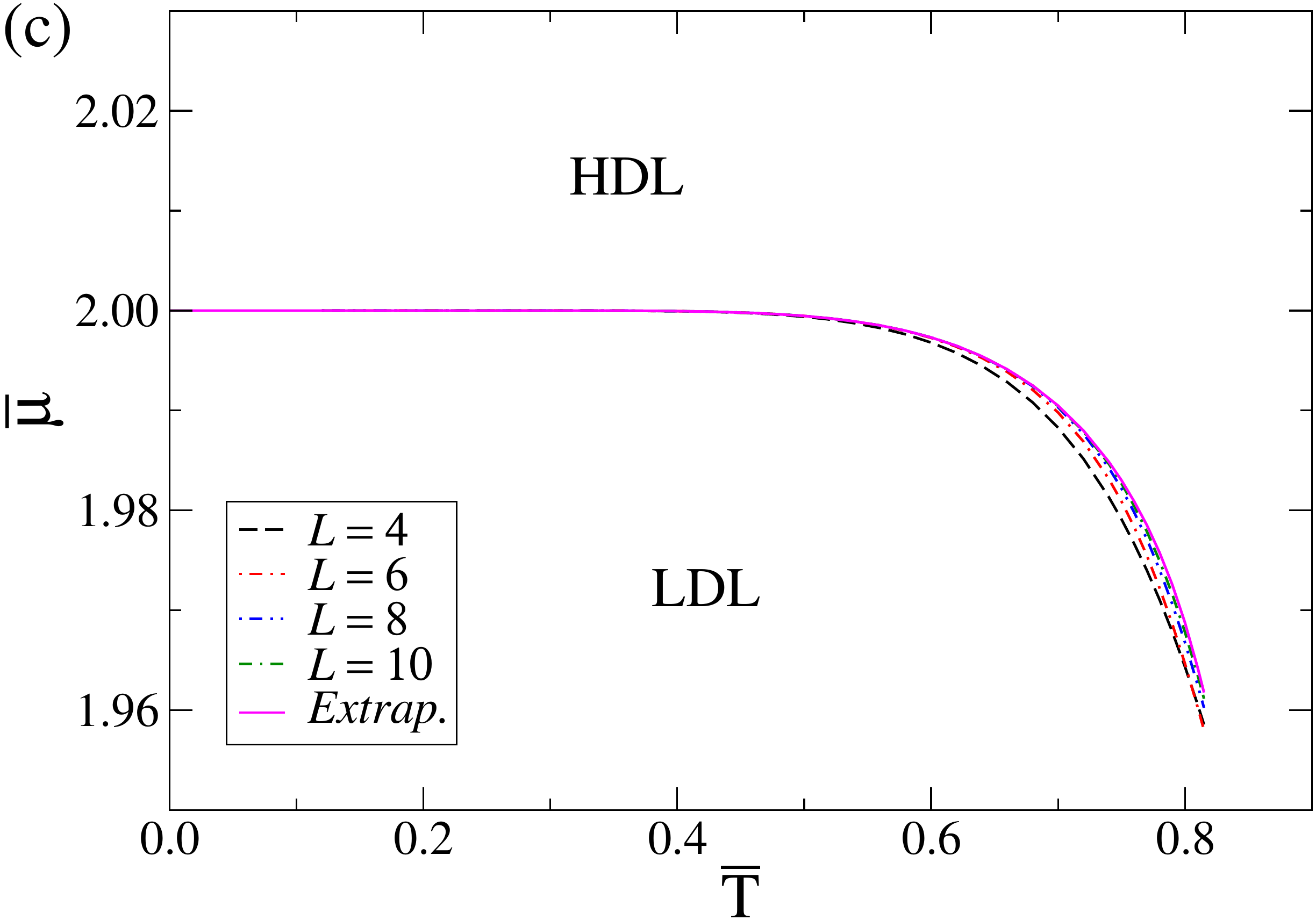}
  \caption{(Color online) a) Estimates of the LDL-HDL coexistence line from the condition $X_3(L-1,\bar{\mu}_c)=X_3(L+1,\bar{\mu}_c)$ for fixed values of $\bar{T}$, for the indicated $L$'s. b) Inverse correlation length $\xi_6^{-1}$ versus $\bar{\mu}$ for $\bar{T}=0.75$ and the indicated $L$'s. c) LDL-HDL coexistence lines obtained from the minima in the curves in (b) and similarly ones for other temperatures. The extrapolated curve [Extrap.], from a 3-pt extrapolation following Eq. \ref{eq:FSScoex}, is also shown.}
  \label{fig:LDLHDL}
\end{figure}

The LDL-HDL transition is the only one for which all previous studies of the ALG model agree; the phases are separated by a coexistence line, which exists in a narrow chemical potential interval near $\bar{\mu}=2$ \cite{Ba07,Ol10,Fu15}. Therefore, in this case, it is convenient to determine the transition lines by varying $\bar{\mu}$ for fixed values of $\bar{T}$. For the strip widths analyzed here, crossings in the curves of $X_k(L;\bar{\mu})$ versus $\bar{\mu}$ for different $L$'s are only observed for $k\geqslant 3$. The curves of $X_1(L;\bar{\mu})$ and $X_2(L;\bar{\mu})$  only exhibit a maximum in the coexistence region. The transition lines obtained from the condition $X_3(L-1,\bar{\mu}_c)=X_3(L+1,\bar{\mu}_c)$ are depicted in Fig. \ref{fig:LDLHDL}(a). The lines for the smallest widths present maxima in the region close to the CEP, which are absent in those for the largest $L$'s. This turns the extrapolation of these lines unreliable in this region, which is the most important part of such curves, since they are always very close to $\bar{\mu}=2$ for low temperatures. 

Motivated by this and also to confirm the first-order nature of the LDL-HDL transition, we obtain an alternative estimate for the LDL-HDL transition line through the correlation length $\xi_k^{-1}(L)$, defined in Eq. \ref{eq:corr_len}, for $k=6$. We remark that, given the four-fold (three-fold) degeneracy of the ground state of the LDL (HDL) phase, a total of seven possible phases (divided into two sets with three and four equivalent ones) coexist at the LDL-HDL transition line. This means that the seven largest eigenvalues (from $\Lambda_0$ to $\Lambda_6$) of the TMs shall be degenerated in the thermodynamic limit ($L \rightarrow \infty$). Hence, curves of $\xi_6^{-1}(L)$ vs $\bar{\mu}$, for a given $\bar{T}$, are expected to display a minimum at the coexistence \cite{Kaski83,Bartelt86,Pla98,Jung17}. As exemplified in Fig. \ref{fig:LDLHDL}(b) for $\bar{T}=0.75$, such behavior is indeed found in these curves, confirming that this is a discontinuous transition. The coexistence lines estimated from the loci of these minima are depicted in Fig. \ref{fig:LDLHDL}(c). It is noteworthy that $\bar{\mu}_c(\bar{T})$ decreases with $L$ in Fig. \ref{fig:LDLHDL}(a), while for these last estimates an increasing behavior is observed in Fig. \ref{fig:LDLHDL}(c).  

In order to extrapolate these lines, we recall that finite-size scaling predicts exponential corrections at the coexistence \cite{fisher71}, such that
\begin{equation}
 \bar{\mu}_c(L) = \bar{\mu}_c^* + a e^{-b L} + \ldots,
 \label{eq:FSScoex}
\end{equation}
for a given $\bar{T}$. For a 3-pt extrapolation this gives $\bar{\mu}_c^* = \frac{\bar{\mu}_c(L+1)\bar{\mu}_c(L-1)-\bar{\mu}_c(L)^2}{\bar{\mu}_c(L+1)+\bar{\mu}_c(L-1)-2\bar{\mu}_c(L)}$. The resulting curve from such a extrapolation, for the three largest $L$'s, is shown in Fig. \ref{fig:LDLHDL}(c).

\subsection{The DF-LDL transition line}
\label{subsec:DF-LDL}
\begin{figure}[t]
  \includegraphics[width=8.5cm]{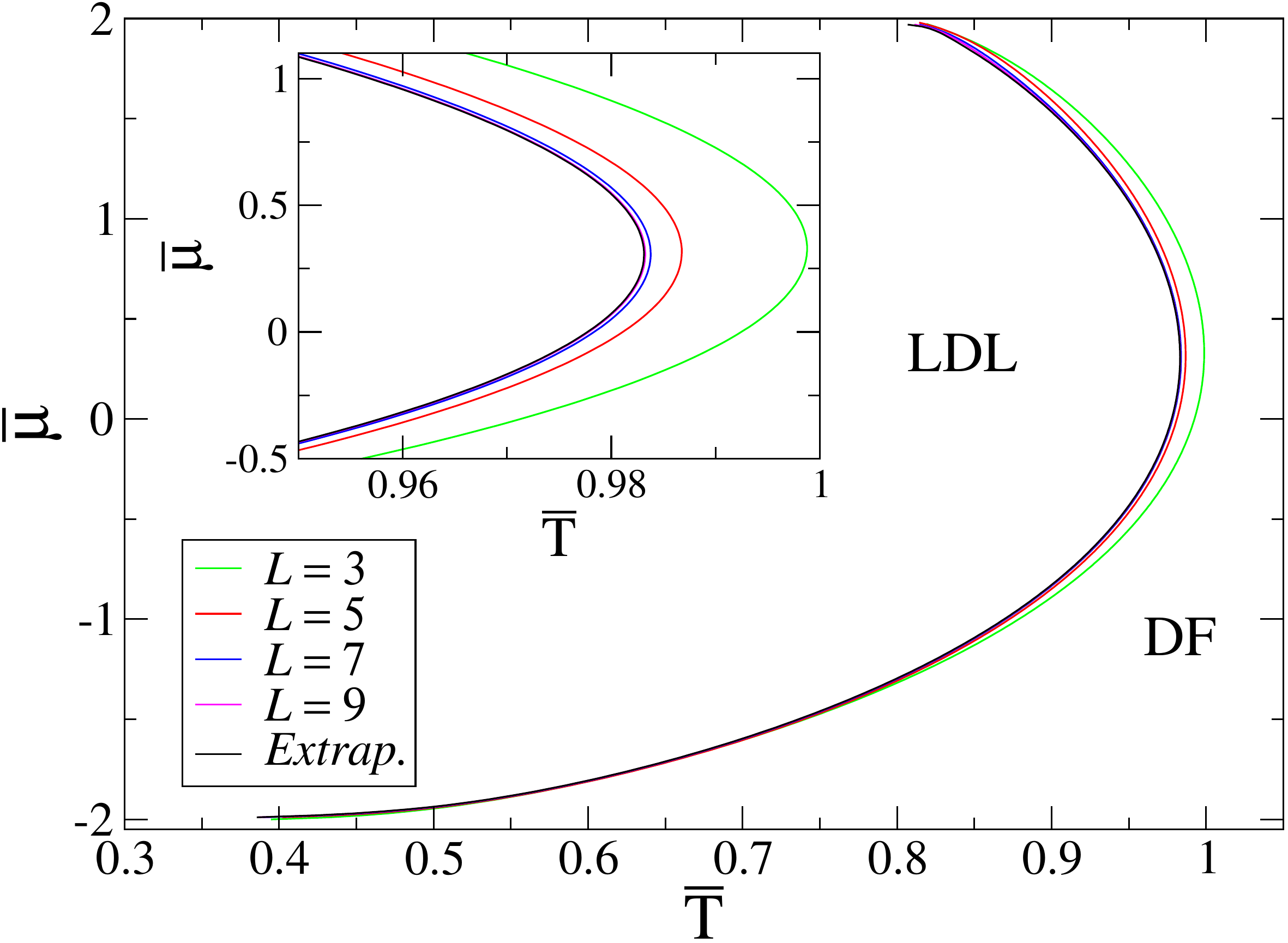}
  \includegraphics[width=8.5cm]{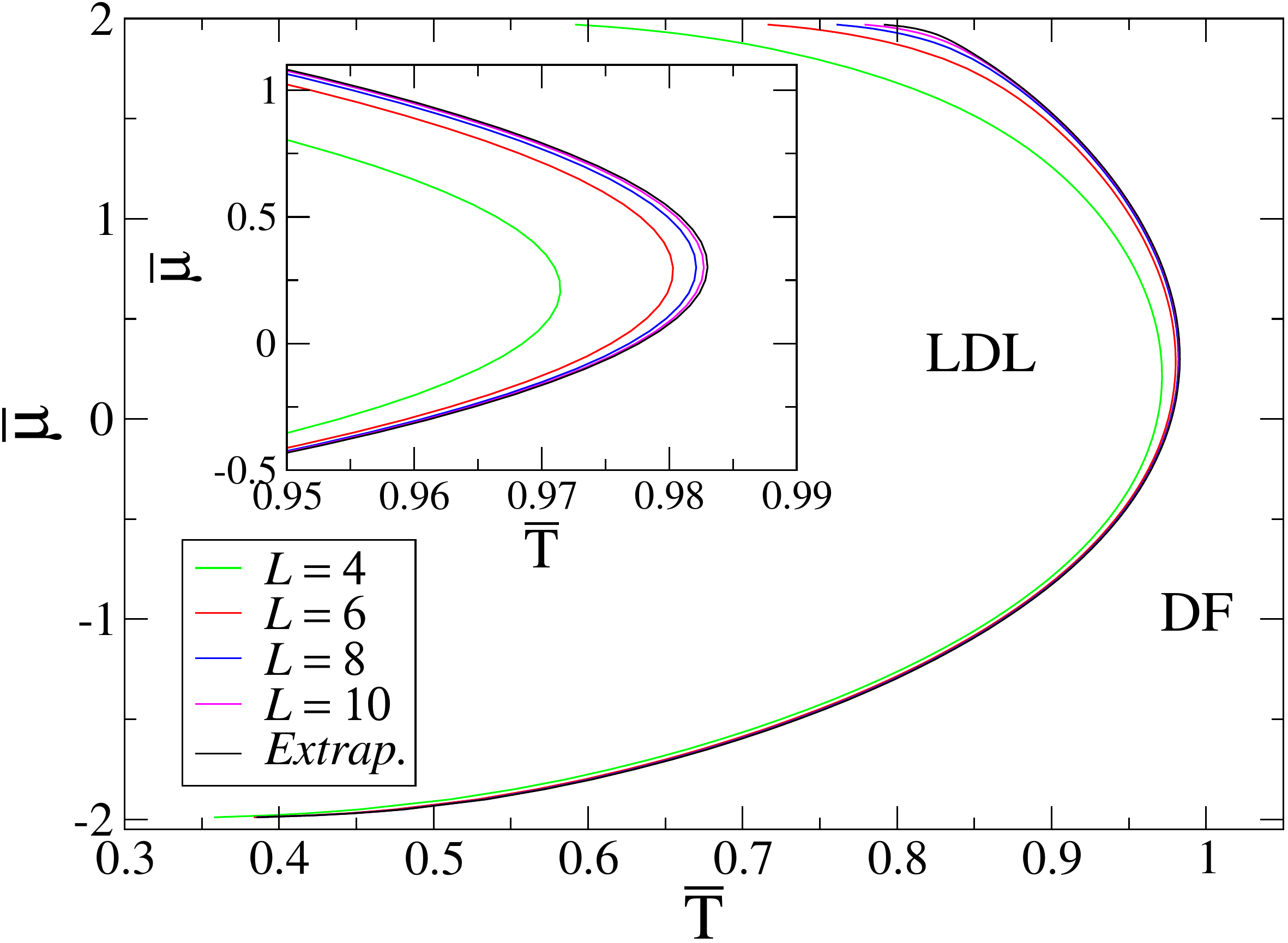}
  \caption{(Color online) Finite-size estimates of the DF-LDL coexistence line: a) from the condition $X_1(L-1,\bar{T}_c)=X_1(L+1,\bar{T}_c)$, for fixed values of $\bar{\mu}$; and b) from the minima in the inverse correlation length $\xi_4^{-1}$. The respective extrapolated curves [Extrap.], from a 3-pt extrapolation following Eq. \ref{eq:FSScoex}, are shown in both panels.}
  \label{fig:GASLDL}
\end{figure}

The transition lines separating the disordered and LDL phases were estimated following the same procedures from the previous subsection, but varying $\bar{T}$ for fixed values of $\bar{\mu}$. The curves obtained from the condition $X_1(L-1,\bar{T}_c) = X_1(L+1,\bar{T}_c)$ are depicted in Fig. \ref{fig:GASLDL}(a), where one sees that $\bar{T}_c$, for a given $\bar{\mu}$ decreases with $L$. Significantly, these lines never cross each other, indicating that no tricritical point exists in the DF-LDL transition, in contrast with the result from MC simulations reported in Ref. \cite{Fu15}. Since one knows that this transition is discontinuous at $\bar{T} = 0$, and for low $\bar{T}$ as well, the absence of the tricritical point implies that the entire transition line is of first-order nature, in agreement with the Husimi lattice solution of the model \cite{Ol10}.

Additional evidence of this is obtained from the curves of $\xi_4^{-1}$ versus $\bar{T}$, for a given $\bar{\mu}$, which have minima hallmarking the coexistence. A behavior analogous to that seen in Fig. \ref{fig:LDLHDL}(b) for $\xi_6^{-1}$ at the LDL-HDL transition. Note that in the case of the DF phase (one ground state) coexisting with the LDL phase (which has four equivalent ground states), minima are indeed expected in $\xi_4^{-1}$. From such minima, we estimate the coexistence lines displayed in Fig. \ref{fig:GASLDL}(b). Interestingly, in this case the values of $\bar{T}_c(\bar{\mu})$ increases with $L$, in opposition to the behavior seen in Fig. \ref{fig:GASLDL}(a). 

Figures \ref{fig:GASLDL}(a) and \ref{fig:GASLDL}(b) present also the asymptotic curves, obtained from 3-pt extrapolations assuming exponential finite-size corrections in $\bar{T}_c(L)$ similarly to Eq. \ref{eq:FSScoex} [i.e., $\bar{T}_c(L) \simeq \bar{T}_c^* + a e^{-b L}$]. The difference between the values of $\bar{T}_c^*(\bar{\mu})$ along these lines is always smaller than $0.5$\% and in the main part of them it is $\lesssim 0.01$\%. The turning point in the coexistence line (i.e., the point where $\bar{T}_c^*$ is maximum) is located at $\bar{T}_c^* \approx 0.983$ and $\bar{\mu}_c^* \approx 0.31$.

The complete phase diagram, presenting all the transition lines found in our MT calculations, will be presented in Sec. \ref{sec:Conc}, where they are compared with the results from the MC simulations discussed in the next section.

\section{Monte Carlo simulations}
\label{sec:MC}

\subsection{Simulation details and quantities of interest}
\label{sec:metsim}

The thermodynamic properties of the symmetric ALG model were also obtained via extensive Monte Carlo~(MC) simulation on the triangular
lattice with periodic boundary conditions. We study system sizes 
$L=32,48,64,96,128$ and $192$ using the standard Metropolis 
algorithm in the grand canonical ensemble. 

Following  references~\cite{Ba07,Fu15}, simulations with fixed chemical potential 
are used 
to locate the DF-HDL line while simulations along isotherms in the $\bar{\mu}-\bar{T}$ plane are used to locate the DF-LDL and LDL-HDL coexistence lines. We study
$\bar{\mu}=1.95,2.3,2.4,2.5, 2.6, 5.0$ and $10$ in the HDL region and
use a temperature interval 
$0.50 \leq\bar T \leq 0.97$ in the LDL region. 

Studies of the HDL region are performed in two steps. First, broad ranges  of temperature are studied in order to identify the region of peaks in the response functions. In this step $50$ independent realizations are used to compute averages. After identifying the region of interest for each system size, additional simulations are performed in a reduced range of temperatures. The number of points around the maximum depends on the system size; for $L=64,128$ we use $60$ points while for larger system sizes $30$ points are sufficient. In this step 100 independent replicas are used. In all simulations we use $10^6$ MC steps for equilibration, followed by $10^5-2\times10^6$ MC steps for data production.

The transition between the HDL and the DF phases can be analyzed using the order parameters, \cite{Sz09,Fu19} 
\begin{equation}\label{eq:defthe} 
\theta(\bar{T},\bar{\mu};L) \equiv \frac{3}{2}\left [ \frac{\max (n_1,n_2,n_3)}{N} - 
\frac{1}{3}\right ],\\
\end{equation}
and
\begin{equation}\label{eq:defq}
    Q(\bar{T},\bar{\mu};L) \equiv \frac{1}{N}\sqrt{n^2_1+n^2_2+n^2_3-n_1n_2-n_1n_3-n_2n_3},\\  
\end{equation}
where $n_k$ denotes the number of molecules in orientational state $k$ and $N = \sum_{k=1}^3 n_k$ is the total number of molecules. In the DF phase, one has $n_1 \approx n_2 \approx n_3$, so that $\theta \rightarrow 0$ and $Q \rightarrow 0$. On the other hand,  $\theta \rightarrow 1$ and $Q \rightarrow 1$ in the HDL phase, where, e.g., $n_1 \approx N$ and $n_2 \approx n_3 \approx 0$. 

To analyze the transitions we also use the susceptibilities related to the order parameters:
\begin{equation}\label{eq:chi}
    \chi_X\prt{\bar T;L} = \frac{L^2}{\bar T}\cch{\aver{X^2} - \aver{X}^2}
\end{equation} 
with $X =\theta$ or $Q$; the specific heat at constant volume~\cite{Al89}
\begin{equation}\label{eq:cv}
      c_V\prt{\bar T;L} = \frac{1}{V\bar T^2}
    \cch{\prt{\aver{E^2}-\aver{E}^2}-
    \frac{\prt{\aver{EN}-\aver{E}\aver{N}}^2 }
    {\aver{N^2}-\aver{N}^2}};
\end{equation}
and the isothermal compressibility
\begin{equation}\label{eq:ic}
      \kappa\prt{\bar T;L} = \frac{V}{k_B\bar T}\prt{
      \frac{\aver{N^2}-\aver{N}^2}{\aver{N}^2}}.
\end{equation}
In Eqs.~\ref{eq:chi}-\ref{eq:ic}, $\aver{\cdot}$ denotes grand canonical averages.


\subsection{Discontinuous transitions}
\begin{figure}
    \centering
    \includegraphics[scale=.5]{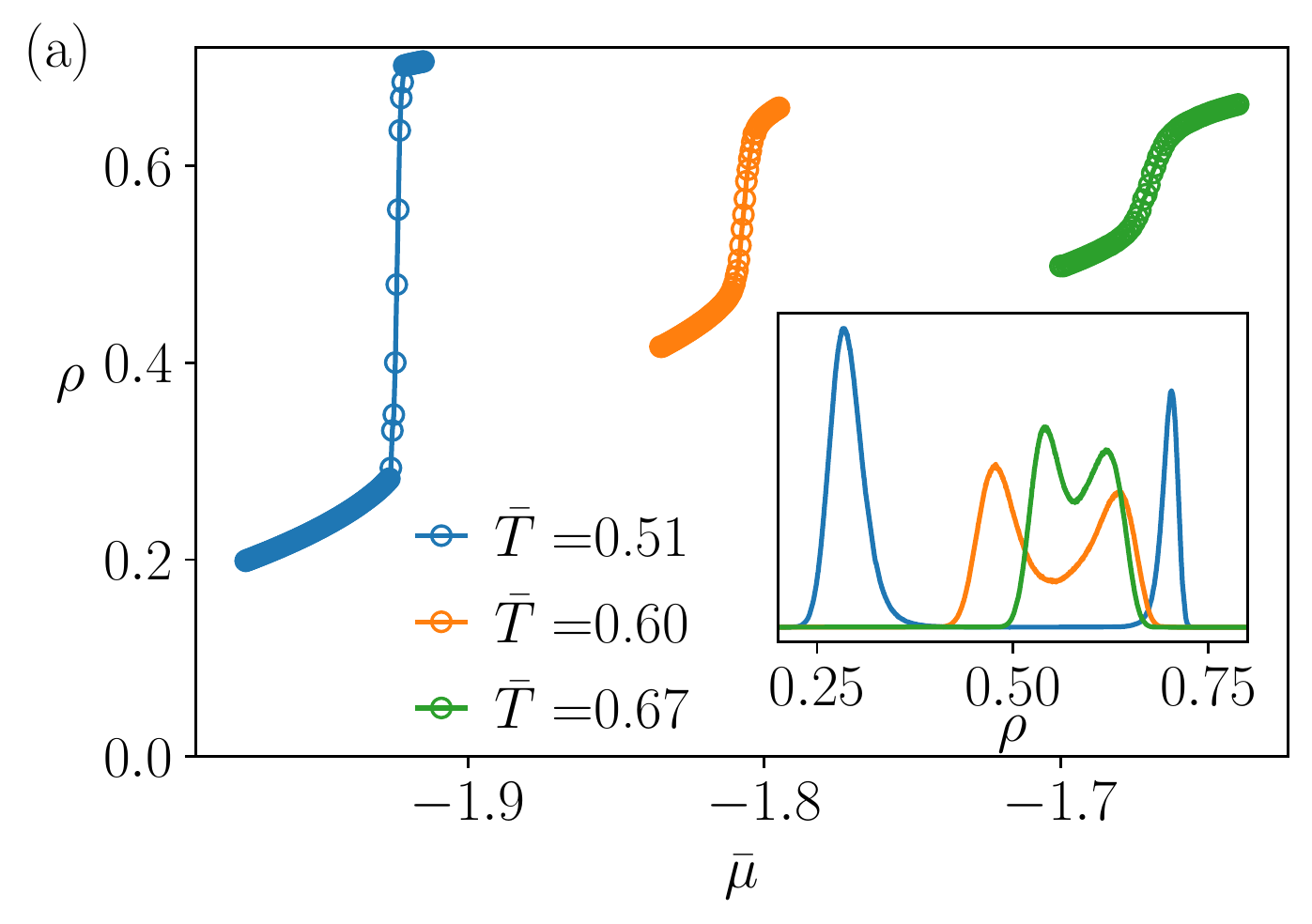}

    \includegraphics[scale=.5]{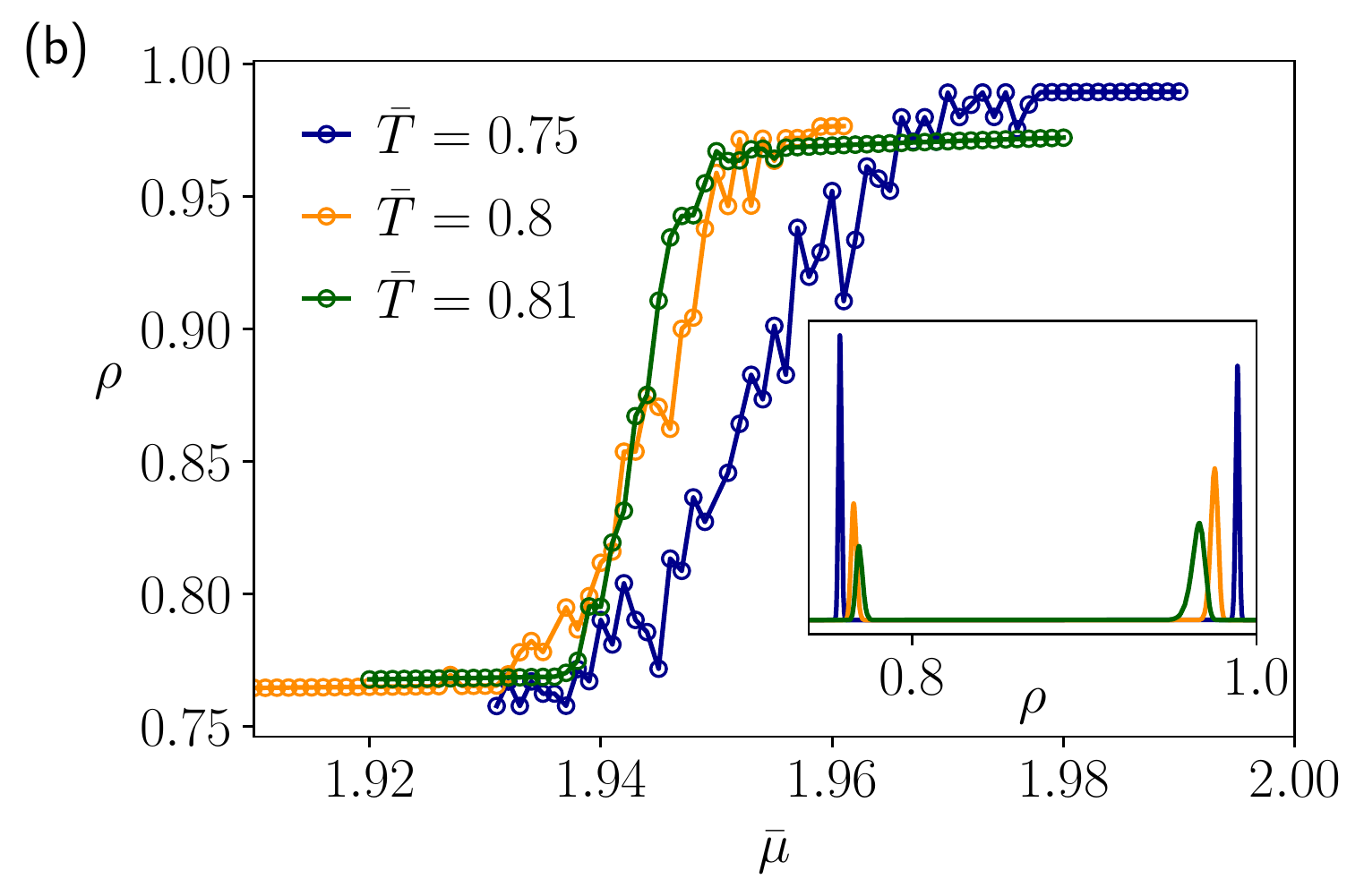}
   
    \caption{Density versus chemical
    potential for distinct temperatures at DF-LDL (a) and LDL-HDL (b) transitions. Insets show histograms of the 
    density for DF-LDL and LDL-HDL near coexistence. Different colors represent
    temperatures as indicated. System size is $L=64$ in (a) and $L=128$ in (b).}
      \label{fig:rhoxmus}
\end{figure}

In this subsection we present the results for DF-LDL
and LDL-HDL transitions. As reported in the Refs.~\cite{Ba07,Fu15} 
the density plays the role of an order parameter for the referred
transitions. Figures \ref{fig:rhoxmus}(a) and (b)  show the density as a function of chemical potential for the  DF-LDL transition  and the LDL-HDL transition, respectively.
The density histograms for each transition are shown in the insets of Figs.~\ref{fig:rhoxmus}(a) and (b). In both cases, the histograms are bimodal indicating a discontinuous transition and confirming the results from the transfer matrix (TM) calculations (see Subsec. \ref{subsec:LDL-HDL} and \ref{subsec:DF-LDL}). For each temperature, we plotted the histogram for the chemical potential that was closest to the coexistence value within our resolution ($\Delta 
\bar{\mu}=10^{-4}$ or larger). We take the coexistence chemical potential as that for which the heights of the two peaks are equal.

\begin{figure}
    \includegraphics[scale=0.5]{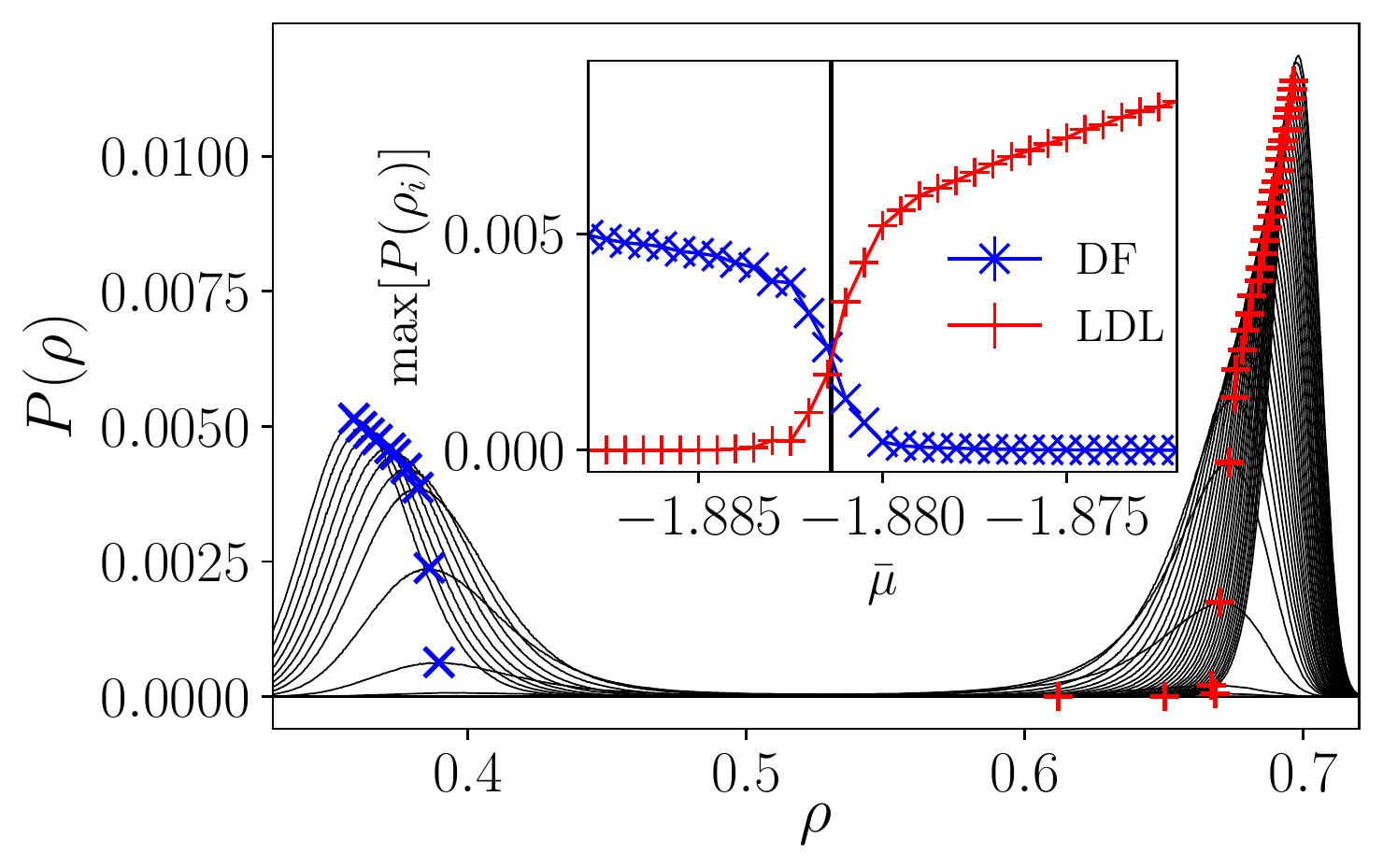}
    \caption{Main panel: thin black lines represent the probability
    distribution of the density $\rho$ for distinct chemical 
    potentials and $\bar T=0.55$. Blue(red) crosses~(pluses) indicate
    the location of the maximum of probability distributions in the DF (LDL) phase.
    Inset: maxima of the
    probability distributions for the disordered phase (blue crosses) and the LDL phase (red pluses) versus chemical potential $\bar{\mu}$. The vertical line indicates the crossing point. System size $L=64$.}
    \label{fig:probrhos}
\end{figure}

The coexistence chemical potential is estimated as follows: 
First we identify the values of the maxima of the probability
distributions, $\max\cch{P\prt{\rho_{DF}}}$ and $\max\cch{P\prt{\rho_{LDL}}}$,
for each chemical potential studied~(see blue crosses and red pluses 
in Fig.~\ref{fig:probrhos}). Subsequently, we plot the 
$\max\cch{P\prt{\rho_{LDL,DF}}}$ as function of chemical potential 
as shown in Fig.~\ref{fig:probrhos} for $\bar T=0.55$. The crossing point of the curves is estimated through a polynomial fit 
in the vicinity of the intersection, using the the reciprocals of the variance ($1/\sigma_i^2$) of the averages computed over the replicas as weights for the polynomial fit of the points. In most cases, 7 points 
around the intersection region are enough to obtain a satisfactory
region of crossing. The order of the polynomial is chosen as the
lowest order that yields random behavior of residuals; a 3rd-order polynomial fit was sufficient in most cases. 
Subsequently,  the error of the fit was computed through the root
mean square error~(RMSE) over the points used in the fit. Finally, the uncertainty in the crossing-point estimate is computed as the square root of the RMSE resulting from each polynomial. This procedure yields the dots in the phase diagram of Fig. \ref{fig:PDmuXT}. 

In the region around $\bar{\mu}=0$, $\bar T>0.95$ the densities $\rho_{LDL}$ and $\rho_{DF}$ are very similar and the coexistence temperature and chemical potential cannot been obtained using the density histogram (see Fig. \ref{fig:PDmuXT}(b)).

For the LDL-HDL transition, the data for the density are very noisy 
for $\bar T\leq 0.75$ (see Fig.~\ref{fig:rhoxmus}(b)) and it is not 
possible to use the procedure described above to determine the 
coexistence chemical potential. For these cases, we adopted the
following procedure: for each point on the curve 
$\max\cch{P\prt{\rho_i}} \times \bar{\mu}$~(see inset of 
Fig.~\ref{fig:probrhos}) we sum a Gaussian random variable proportional to 
the error bar of that point. We then fit
the curves and estimate the crossing point. An average is computed over $10^4$ realizations of this procedure and the error is propagated as described above. The values obtained using this strategy are the triangles in Fig. {\ref{fig:PDmuXT}} and are in good agreement with the TM results of Subsec. \ref{subsec:LDL-HDL}. 

The highest temperature for which we observe LDL-HDL coexistence is $\bar T=0.816$, $\bar \mu=1.9498(7)$, for the next temperature studied, $\bar T=0.817$, we observe LDL-DF coexistence and then a continuous DF-HDL transition. We therefore take this  $\bar T$, $\bar \mu$ pair as an estimate of the critical end point (CEP) at which the LDL-HDL line meets the DF-HDL and the DF-LDL lines (see Subsec. \ref{subsec:DF-HDL}). This value, $\bar T_{cep}=0.816$, $\bar \mu_{cep}=1.9498(7)$, is in excellent agreement with the estimate from TM calculations, $\bar T_{cep,TM}=0.815(1)$, $\bar \mu_{cep,TM}=1.9597(8)$. Our estimate for $\bar T_{cep,MC}$ and $\bar \mu_{cep,TM}$ is smaller than the values reported in previous MC simulations \cite{Ba07,Fu15} for $c_2$ and $tc_2$ (see Fig. \ref{fig:reffigs}).

\subsection{Continuous transitions}

To determine the critical temperatures we use finite-size 
scaling~(FSS)\cite{Al89} analysis of the susceptibilities 
$\chi_X\prt{\bar T;L}$  
of the order parameters  $X=\theta$ and $Q$ (Eq. \ref{eq:chi}), the specific heat at constant volume $c_V\prt{\bar T;L}$ (Eq. \ref{eq:cv}), and isothermal compressibility (Eq. \ref{eq:ic}).

\begin{figure*}[!htb]
    \includegraphics[scale=.6]{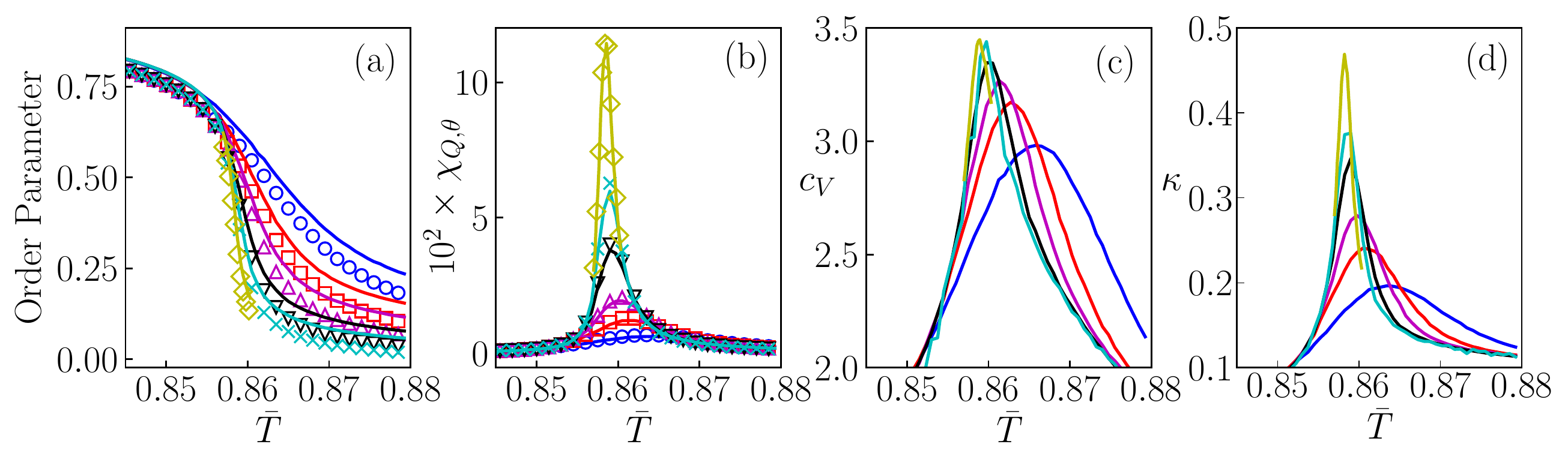}
 
    \caption{ (a) Order parameters $Q$ and $\theta$, (b) susceptibilities $\chi_Q$ and $\chi_\theta$, (c) specific heat $c_V$, and (d) isothermal compresibility for $\bar \mu=2.3$. Blue, red,  magenta, black, cyan, and yellow
    correspond to  system sizes $L=32,48,64,96,128$ and 
    $192$.  The order parameter $Q$ and its susceptibility 
    $\chi_Q$ are represented by solid lines, $\theta$  
    and $\chi_\theta$ by symbols.}
    \label{fig:propsxT}
\end{figure*}

Figure \ref{fig:propsxT} shows simulation results for the order parameters, $Q$ and $\theta$; susceptibilities, 
$\chi_Q$ and $\chi_\theta$, specific heat $c_V$, and isothermal compressibility $\kappa$ for chemical potential
$\bar{\mu}=2.3$. For clarity, error 
bars are not shown in the figure. The 
susceptibility, specific heat and isothermal compressibility exhibit signatures of a continuous transition. In all cases the peaks increase systematically with system size. Similar results are obtained for $\bar{\mu}=2.4,2.5,2.6,5.0$, and 10. As the chemical potential is increased, the maxima shift to higher temperatures, approaching the value  $\bar T^*_{c,\bar{\mu}\to\infty}=0.9526$ reported in Ref. \cite{Fu19}. This behavior confirms that the DF-HDL line is continuous, in agreement with the TM calculations from Subsec. \ref{subsec:DF-HDL}.

Using the results for $\chi_Q,\chi_\theta,c_V$ and $\kappa$ we 
estimate the critical temperature $\bar T_c^*$. Initially, we estimate the size dependent {\it pseudocritical temperature} $\bar T_c$ through a polynomial fit to the data near the maximum, as described in the preceding subsection.

\begin{figure}[!htb]
    \includegraphics[scale=.5]{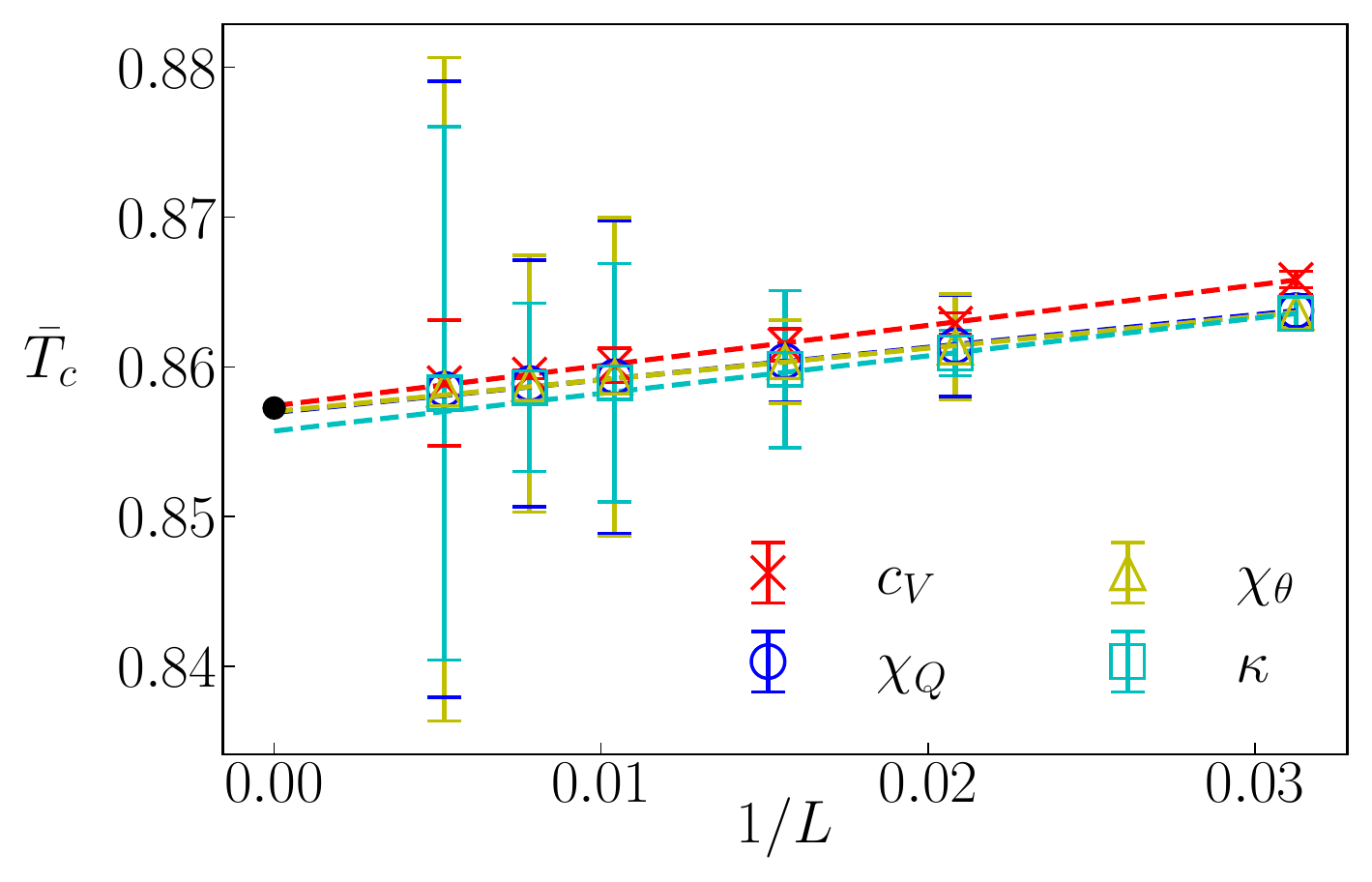}
    \caption{Pseudocritical temperatures associated with, $c_v$, $\chi_Q$, $\chi_{\theta}$ and $\kappa$, with symbols as indicated, versus $1/L$ for  $\bar{\mu}=2.3$. Dash lines are linear fits to the data and the filled black dot indicates the global estimate for the critical temperature $\bar T^*_c$.}
    \label{fig:FSS}
\end{figure}

The pseudocritical temperatures $\bar T_c$ for  $\bar{\mu}=2.3$ are plotted versus $1/L$ in Fig.~\ref{fig:FSS}. Similar results were obtained for $\bar{\mu}=2.4,2.5,2.6,5.0$ and $10$. The set of pseudocritical temperatures
appear to converge as $L\to\infty$ for all the studied values of $\bar \mu$. The global estimate of $\bar T^*_c$ was obtained though a weighted average with
weight $1/\sigma^2$, where $\sigma$ represents the uncertainty
of each quantity. The final estimates of $\bar T^*_c$ are the squares in the phase diagram Fig. \ref{fig:PDmuXT}.

\subsection{Critical exponents}

\begin{figure}[!htb]
  \includegraphics[scale=0.45]{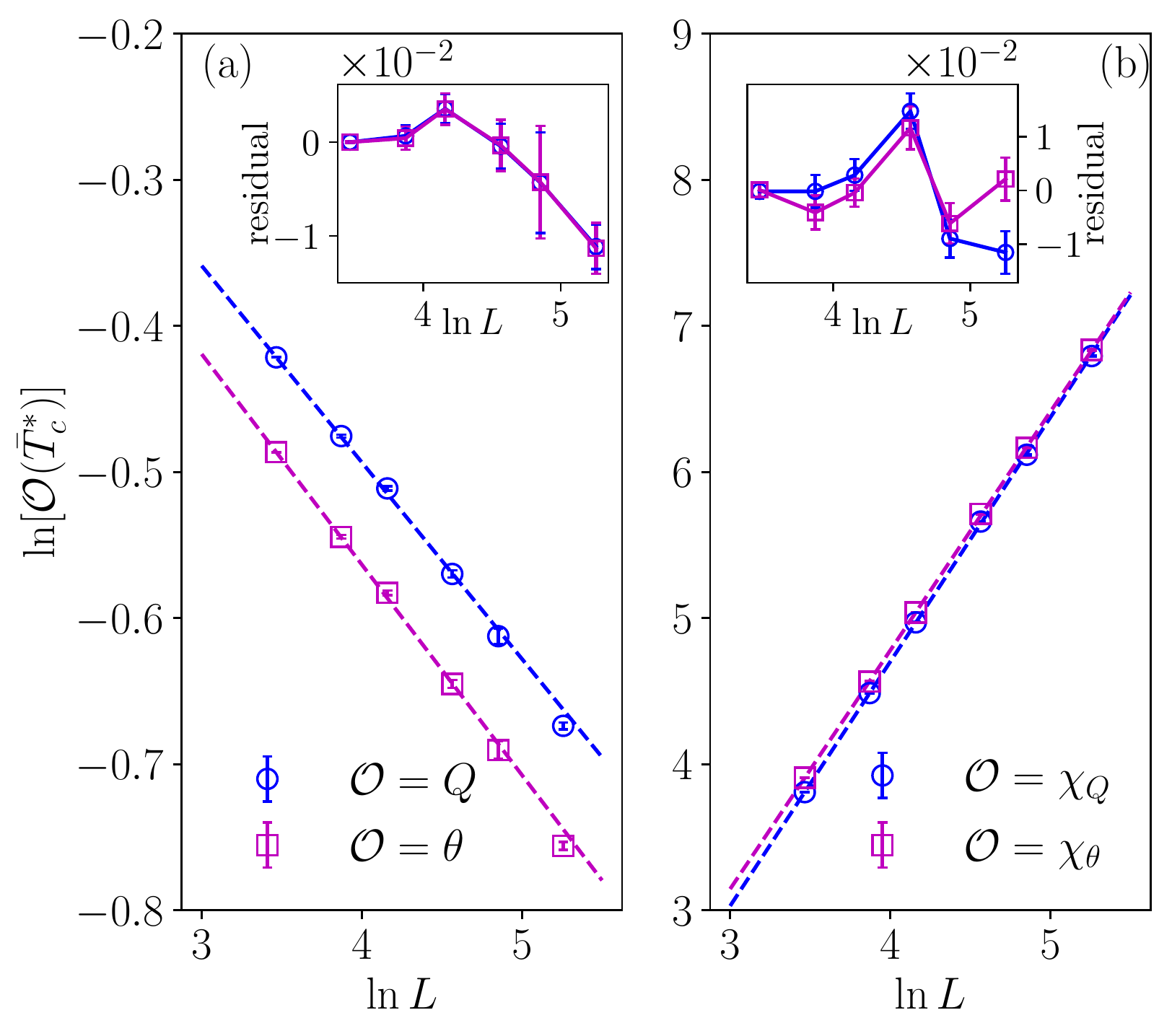}
        \caption{(a) Order parameter $Q$($\theta$) versus system size for $\bar \mu=2.3$. The circles (squares) are the simulation data; dashed lines are least-squares linear fits of the data. Inset shows the residuals of the fits. (b) Analogous plots for the susceptibilities $\chi_Q$($\chi_\theta$). }
  \label{fig:bn-gn}
\end{figure}

For $\bar \mu=2.3$ and $10$, we estimate the critical- exponent ratios $\beta/\nu$ and $\gamma/\nu$ from fits to the data for $Q(\bar T^*_c)$, $\theta(\bar T^*_c)$, $\chi_Q(\bar T^*_c)$, and $\chi_\theta(\bar T^*_c)$ versus system size (see Fig. \ref{fig:bn-gn}).
From the linear fits, and including the effect of the uncertainty in $\bar T^*_c$, we obtained the estimates in Table \ref{tab:CE}. The values for these ratios for the two-dimensional three-state Potts model are $\beta_{Potts}/\nu_{Potts}=2/15=0.1333\cdots$ and $\gamma_{Potts}/\nu_{Potts}=26/15=1.7333\cdots$. In the worst cases, the discrepancy is about $6\%$ for $\mu=2.3$ and $8\%$ for $\mu=10$ . 

\begin{figure}[!htb]
  \includegraphics[scale=0.45]{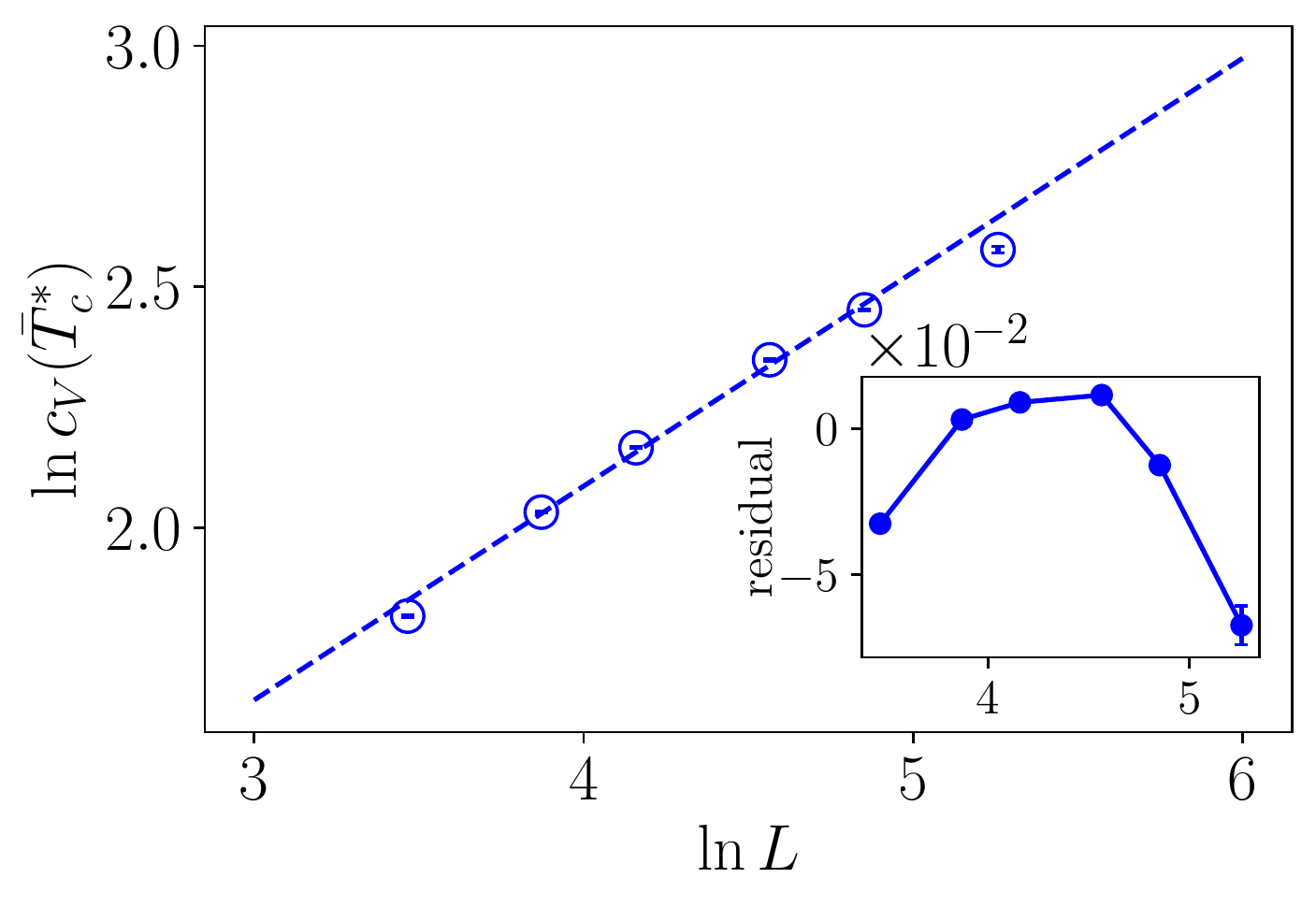}
        \caption{Specific heat $c_V$ versus system size at critical temperature for $\bar \mu=10.0$ The circles are the simulation data; the dashed lines is a least-squares linear fit of the data. Inset shows the residuals of the fit.}
  \label{fig:cvFSS}
\end{figure}

Figure \ref{fig:cvFSS} shows the linear fit for $\ln c_v(\bar T_c)$  for $\mu=10.0$, yielding the estimate $\alpha/\nu=0.44(1)$, i.e., a discrepancy with the ratio for the two-dimensional, three-state Potts model $\alpha/\nu_{Potts}=0.4$ of 10\%. For $\bar \mu=2.3$, the discrepancy in this ratio is $73\%$. We believe this huge difference is likely due to finite-size corrections that are more important close to the CEP ($\bar\mu_{cep}=1.9498$) where the DF-HDL line encounters the LDL-HDL and DF-LDL lines.

\begin{table}[!htb]

    \centering
    \begin{tabular}{cccccc}
    \hline \hline
$\mu$ & $\beta_{Q}/\nu$ & $\beta_{\theta}/\nu$ & $\gamma_{Q}/\nu$
& $\gamma_{\theta}/\nu$ & $\alpha/\nu$\\ \hline
2.3 & 0.126(7) & 0.136(8) & 1.66(4) & 1.63(4) & 0.11(1)  \\
10.0 & 0.126(7) & 0.122(7) & 1.75(5)& 1.75(4) & 0.44(1)\\ \hline \hline
\end{tabular}
\caption{Estimates of the ratios $\beta/\nu$, $\gamma/\nu$ and $\alpha/\nu$ for chemical potentials $\bar \mu=2.3$ and 10.}
\label{tab:CE}
\end{table}

\section{Discussion}
\label{sec:Conc}

\begin{figure}[!htb]
  \includegraphics[scale=0.55]{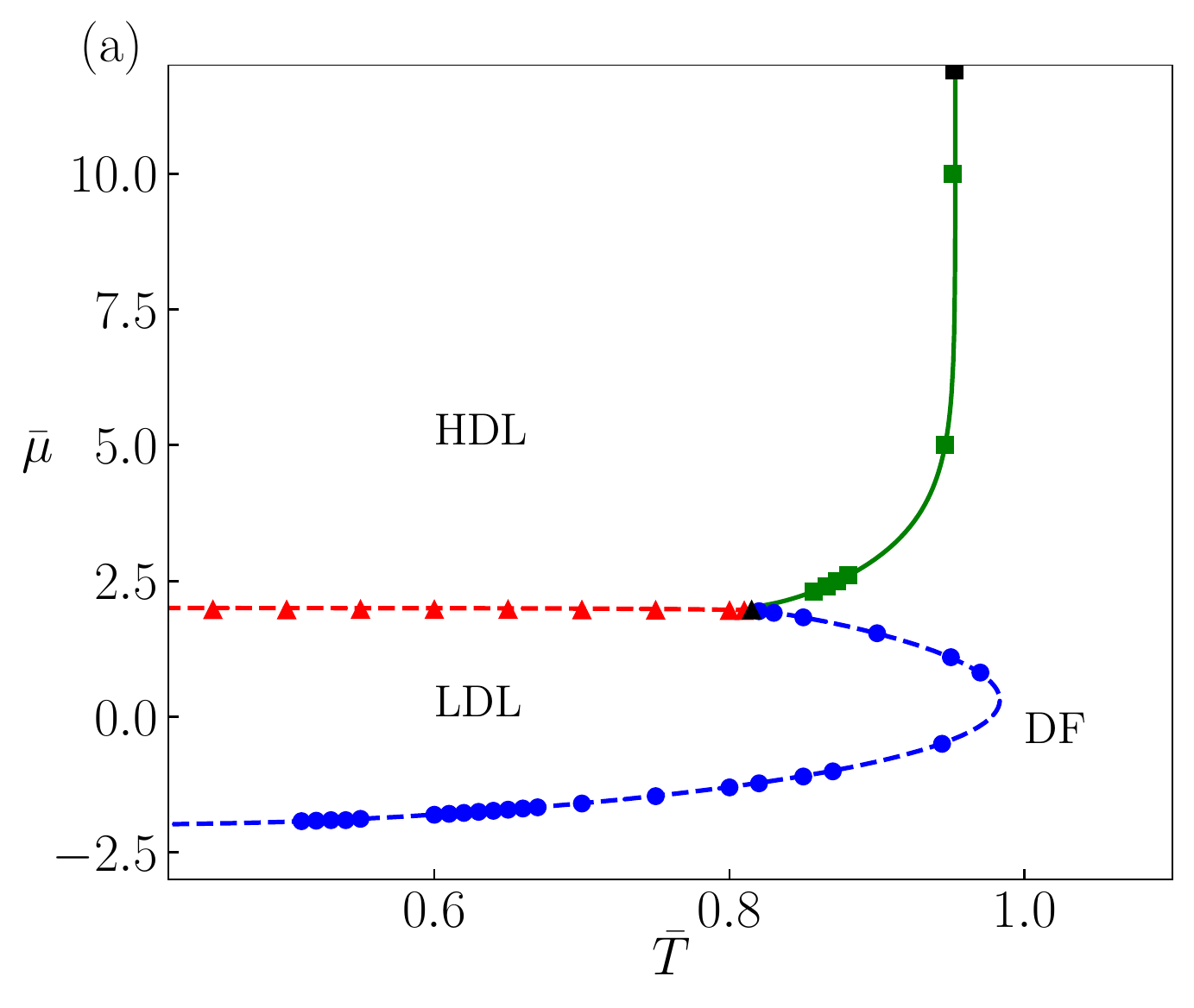}
   \includegraphics[scale=0.55]{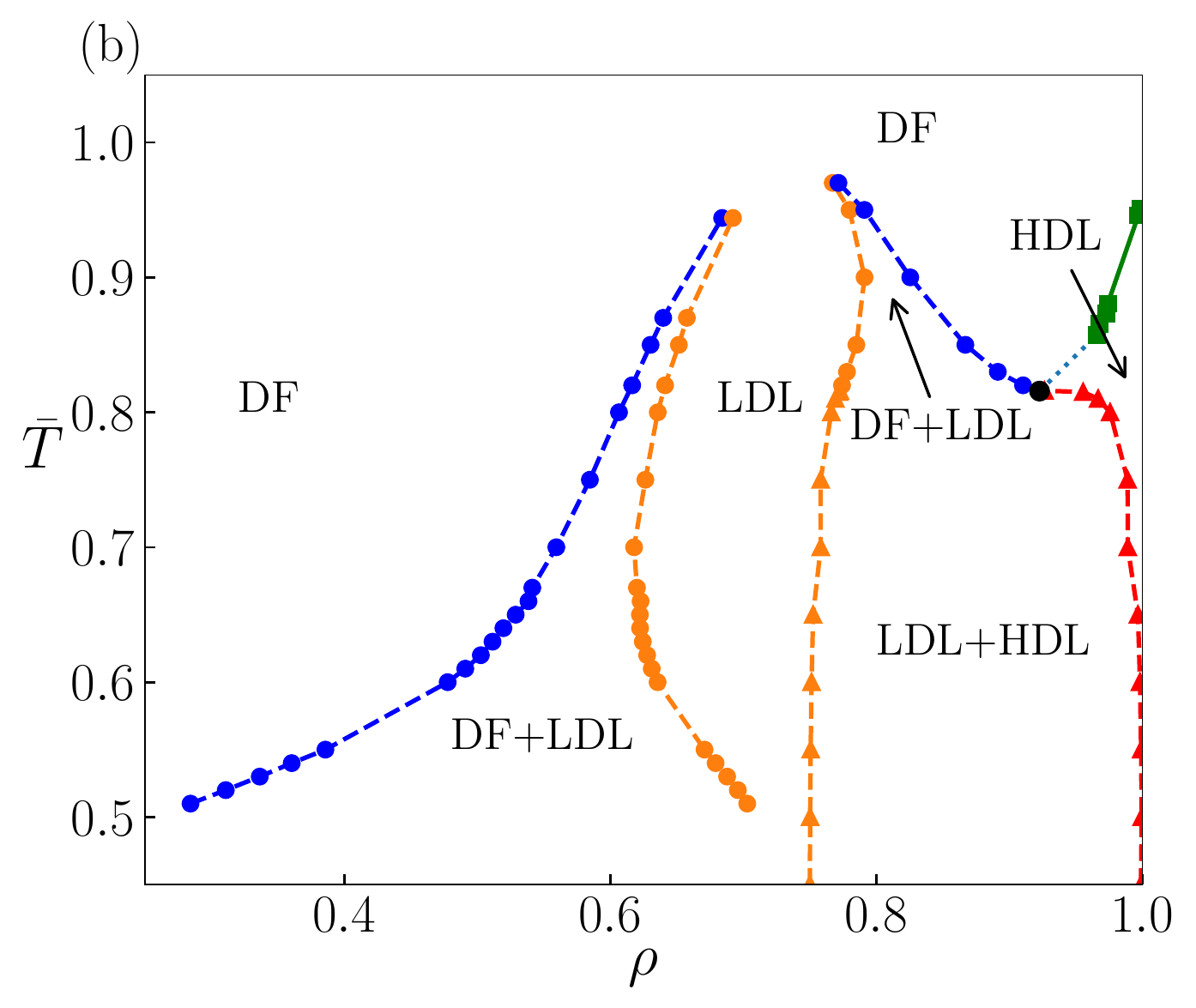}
        \caption{Phase diagram for the symmetric ALG model in the $\bar{\mu} \times \bar T$ plane (a) and the $\bar T\times \rho$ plane (b). Symbols are results from Monte Carlo (MC) simulations.  Dashed lines represent discontinuous transitions and solid ones continuous transitions. Only in (a) Lines are results obtained using Transfer Matrix (TM) calculations. The black triangle marks the location of the critical end point (CEP) estimated using TM, the black square marks the location of the critical point in the limit $\bar \mu \to \infty$ obtained via MC simulation in Ref. \cite{Fu19} and the black dot marks the density at $\bar \mu=1.95$ and $L=128$, very close to our estimate of $\bar \mu_{cep}=1.9498(6)$. The green dotted line 
         is a conjecture as to how to critical line meets the CEP.}
  \label{fig:PDmuXT}
\end{figure}

Figure \ref{fig:PDmuXT}(a) shows the phase diagram of the symmetric ALG model in the $\bar{T}-\bar{\mu}$ plane, summarizing the  results reported above. 
The predictions of the TM analysis are in very good
quantitative agreement with MC simulation. In fact, the maximum difference between the estimates from each method is smaller than 2\% along the LDL-HDL coexistence line and, notably, is at most 0.1\% for the DF-HDL and DF-LDL lines. The natures of the transitions are also the same in both cases. This remarkable agreement confirms that we are accessing the correct thermodynamic behavior of the model. 

Our results demonstrate that the DF-LDL transition is always discontinuous, without any critical or multicritical point. This is confirmed by the absence of a crossing in the pseudocritical transition lines from the TM analysis, by the bimodal histograms in MC simulations and also in the phase diagram of temperature versus total density ($\bar{T}-\rho$), depicted in Fig. \ref{fig:PDmuXT}(b). This diagram makes it clear that the densities are discontinuous at the DF-LDL transition line up to temperatures very close to the turning point (in the $\bar{\mu}-\bar{T}$ diagram), which is located at $\bar{T} \approx 0.983$. It is, however, difficult to estimate the densities accurately very near this point, since $\rho_{DF}$ becomes very similar to $\rho_{LDL}$. An analogous behavior was found in the Husimi lattice solution of this model \cite{Ol10} and indeed our ($\bar{T}-\rho$) diagram is very similar to the one found there [see Fig. 4(a) of Ref. \cite{Ol10}]. Importantly, these results provide compelling evidence that the entire DF-LDL transition line is discontinuous, ruling out a critical point at $\bar{T} \approx 0.55$ or a tricritical point at $\bar{T} \approx 0.65$, as claimed in previous MC analyses in references \cite{Ba07} and \cite{Fu15}, respectively.

Also in contrast with earlier simulations, we find a critical line separating the DF and HDL phases. This line meets the LDL-HDL and DF-LDL coexistence lines at a critical end point (CEP). This shows that the point where the LDL-HDL coexistence line ends --- reported in literature as critical \cite{Ba07}, triple \cite{Ol10} and tricritical \cite{Fu15} --- is actually a CEP. Therefore, although this ALG model has an interesting phase behavior, beyond displaying some thermodynamic anomalies, it is not a good model for liquid water, as initially suggested, because it has neither a liquid-gas coexistence line ending at a critical point, nor a LLCP. In light of the discussion in the Introduction, it seems that absence of a LLCP is a general feature of lattice models with attractive orientational interactions competing with ``van der Waals" repulsion between first neighbor molecules. 

Although determining the universality class of the continuous transitions displayed by models for water is key to establish a full connection with the actual behavior of water, this has not been widely investigated, perhaps because the very nature of these transitions are the subject of controversy. A recent account of these critical properties for realistic models can be found in Ref. \cite{De20}. For lattice models, some examples include the debate in the literature on the critical exponents of the liquid-liquid transition in the classic Bell-Lavis model \cite{Fi09,Si14}, and our previous study of the present ALG model in the fully occupied limit, where the DF-HDL transition was found to be in the 3-state Potts class. The central charge obtained here, deviating by $\sim 1$\% from the Potts value, gives additional evidence of this. The scaled gaps, on the other hand, present a large deviation from those of the 3-state Potts class. The analysis of the central charge for finite chemical potentials indicates that it converges to the same value along the entire DF-HDL critical line. Nevertheless, finite-size corrections become very strong as $\bar{\mu}$ approaches $\bar{\mu}_{CEP} = 1.9597(8)$. This certainly explains why some critical exponent ratios estimated in simulations for $\bar{\mu} = 2.3$ exhibit considerable deviations from the Potts values, while for $\bar{\mu} = 10$ the deviations are much smaller. Despite this, the overall picture suggests that this order-disorder transition is always in the 3-state Potts class, as is expected from its three-fold ($S_3$) symmetry breaking.
\section*{Acknowledgments}
 Authors acknowledge financial support from CAPES Brazil and the National Institute of Science and Technology for Complex Systems of Brazil.

\section*{DATA AVAILABILITY}
The data that support the findings of this study are available from the corresponding author upon reasonable request.

\bibliography{references.bib}

\end{document}